\documentclass[12pt]{article}
\pdfoutput=1 
\usepackage{amssymb}
\usepackage{epsfig}
\usepackage{bbm}
\usepackage{bbold}
\usepackage{graphicx}

\def\L{\mathcal L}
\def\e{\varepsilon}

\newcommand{\wt}{\widetilde}

\begin{document}

\def\a{\alpha}
\def\b{\beta}
\def\c{\chi}
\def\d{\delta}
\def\e{\epsilon}
\def\f{\phi}
\def\g{\gamma}
\def\h{\eta}
\def\i{\iota}
\def\j{\psi}
\def\k{\kappa}
\def\l{\lambda}
\def\m{\mu}
\def\n{\nu}
\def\o{\omega}
\def\p{\pi}
\def\q{\theta}
\def\r{\rho}
\def\s{\sigma}
\def\t{\tau}
\def\u{\upsilon}
\def\x{\xi}
\def\z{\zeta}
\def\D{\Delta}
\def\F{\Phi}
\def\G{\Gamma}
\def\J{\Psi}
\def\L{\Lambda}
\def\O{\Omega}
\def\P{\Pi}
\def\Q{\Theta}
\def\S{\Sigma}
\def\U{\Upsilon}
\def\X{\Xi}

%Varletters
\def\ve{\varepsilon}
\def\vf{\varphi}
\def\vr{\varrho}
\def\vs{\varsigma}
\def\vq{\vartheta}

\def\dg{\dagger}                                     % hermitian conjugate
\def\ddg{\ddagger}                                   % double dagger
\def\wt#1{\widetilde{#1}}                    % big tilde
\def\mt{\widetilde{m}_1}
\def\mti{\widetilde{m}_i}
\def\mtj{\widetilde{m}_j}
\def\rt{\widetilde{r}_1}
\def\mtt{\widetilde{m}_2}
\def\mttt{\widetilde{m}_3}
\def\rtt{\widetilde{r}_2}
\def\mb{\overline{m}}
\def\VEV#1{\left\langle #1\right\rangle}        % < >
\def\be{\begin{equation}}
\def\ee{\end{equation}}
\def\ds{\displaystyle}
\def\ra{\rightarrow}

\def\bea{\begin{eqnarray}}
\def\eea{\end{eqnarray}}
\def\NO{\nonumber}
\def\Bar#1{\overline{#1}}

% Journal abbreviations (preprints)

\def\pl#1#2#3{Phys.~Lett.~{\bf B {#1}} ({#2}) #3}
\def\np#1#2#3{Nucl.~Phys.~{\bf B {#1}} ({#2}) #3}
\def\prl#1#2#3{Phys.~Rev.~Lett.~{\bf #1} ({#2}) #3}
\def\pr#1#2#3{Phys.~Rev.~{\bf D {#1}} ({#2}) #3}
\def\zp#1#2#3{Z.~Phys.~{\bf C {#1}} ({#2}) #3}
\def\cqg#1#2#3{Class.~and Quantum Grav.~{\bf {#1}} ({#2}) #3}
\def\cmp#1#2#3{Commun.~Math.~Phys.~{\bf {#1}} ({#2}) #3}
\def\jmp#1#2#3{J.~Math.~Phys.~{\bf {#1}} ({#2}) #3}
\def\ap#1#2#3{Ann.~of Phys.~{\bf {#1}} ({#2}) #3}
\def\prep#1#2#3{Phys.~Rep.~{\bf {#1}C} ({#2}) #3}
\def\ptp#1#2#3{Progr.~Theor.~Phys.~{\bf {#1}} ({#2}) #3}
\def\ijmp#1#2#3{Int.~J.~Mod.~Phys.~{\bf A {#1}} ({#2}) #3}
\def\mpl#1#2#3{Mod.~Phys.~Lett.~{\bf A {#1}} ({#2}) #3}
\def\nc#1#2#3{Nuovo Cim.~{\bf {#1}} ({#2}) #3}
\def\ibid#1#2#3{{\it ibid.}~{\bf {#1}} ({#2}) #3}

\title{\bf
Leptogenesis with heavy neutrino flavours:
from density matrix to Boltzmann equations}

\author
{\Large Steve Blanchet$^{a}$,  Pasquale Di Bari$^{b}$,  David~A.~Jones$^{b}$, Luca Marzola$^{b}$
\\
$^a$
{\it\small Institut de Th\'eorie des Ph\'enom\`enes Physiques,}
{\it \small \'Ecole Polytechnique F\'ed\'erale de Lausanne, }\\
{\it \small CH-1015 Lausanne, Switzerland} \\
$^b$
{\it\small School of Physics and Astronomy},
{\it\small University of Southampton,}
{\it\small  Southampton, SO17 1BJ, U.K.}
}

\maketitle \thispagestyle{empty}

\vspace{-8mm}
%\centerline{\date{\today}}

\begin{abstract}
Leptogenesis with heavy neutrino flavours is discussed within a density matrix
formalism. We write the density matrix equation, describing the generation of
the matter-antimatter asymmetry,
for an arbitrary choice of the right-handed (RH) neutrino masses.
For  hierarchical RH neutrino masses lying in the fully flavoured regimes,
this reduces to multiple-stage Boltzmann equations. In this case
we recover and extend results previously derived within  a  quantum state collapse description.
We confirm the generic existence of phantom terms. However, taking into account the
effect of gauge interactions, we show that  they are
washed out at the production with a wash-out rate that is halved compared to that one acting
on the total asymmetry. In the $N_1$-dominated scenario they cancel without
contributing to the final baryon asymmetry. In other scenarios they do not in general
and they have to be taken into account.
We also confirm that there is a (orthogonal) component in the asymmetry produced by
the heavier RH neutrinos which completely escapes the washout from the lighter RH neutrinos and show
that phantom terms additionally contribute to it. The other (parallel) component is washed
out with the usual exponential factor, even for  weak washout.
Finally, as an illustration, we study the two RH neutrino model in the light of
the above findings, showing that phantom terms can contribute
to the final asymmetry also in this case.
\end{abstract}

\newpage

%%%%%%%%%%%%%
\section{Introduction}
%%%%%%%%%%%%%

Leptogenesis \cite{fy} is a direct cosmological application of the see-saw mechanism
\cite{seesaw}
for the explanation of neutrino masses and mixing and it  realises a highly non trivial link
between cosmology and neutrino physics.
The discovery of neutrino masses and mixing in neutrino oscillation experiments \cite{neuosc}
has drawn great attention on leptogenesis that became one of the most attractive models
of baryogenesis for the  explanation of the matter-antimatter  asymmetry of the Universe.

In most cases, classical Boltzmann equations are sufficient for the calculation
of the final asymmetry \cite{DL2,early,bcst,cmb,giudice,pedestrians}.
However, when lepton flavour effects are taken into account \cite{bcst,nardi,abada},
different sets of classical Boltzmann equations apply
depending whether the asymmetry is generated in the one-flavour regime, when the mass
of the decaying RH neutrinos  $M_i$ is much above $10^{12}$ GeV, in the two-flavour regime, for
$10^{12}\,{\rm GeV}\gg M_i \gg 10^9\,{\rm GeV}$, or in the  three-flavour regime for $M_i \ll 10^9\,{\rm GeV}$.
Moreover classical Boltzmann equations fail in
reproducing the correct result in the transition regimes for $M_i \sim 10^{9}\,{\rm GeV}$
and for $M_i \sim 10^{12}\,{\rm GeV}$.  However, in the case that
only the lightest RH neutrino species
is assumed to be responsible for the generation of the asymmetry,  classical Boltzmann equations
provide quite a convenient description in phenomenological investigations, since
the final asymmetry can be expressed in terms of simple analytical expressions that well approximate
the numerical solutions \cite{pedestrians,flavorlep}.

On the other hand,  the contribution
from heavier RH neutrinos can also relevantly contribute to
the final asymmetry (heavy neutrino flavour effects) and has,
therefore, consistently to be taken into account in general \cite{geometry}.
When lepton flavour effects are  jointly considered \cite{bcst,vives},
a reliable calculation of the asymmetry cannot neglect the contribution
from the heavier RH neutrinos even in the two RH neutrino model \cite{2RHneutrino}
usually considered as  a paradigmatic case for the validity of the traditional $N_1$-dominated scenario.
It has  also been shown that a successful $N_2$-dominated scenario is naturally realised
in the interesting class of $SO(10)$ inspired models \cite{SO10lep}.

When heavier RH neutrinos are included, one has to distinguish quite a large number of possible mass patterns
with different corresponding sets of classical Boltzmann equations for the calculation of the final asymmetry.
For example, in the typical case of three RH neutrinos one has ten different possible mass patterns \cite{problem} shown in Fig.~1.
\begin{figure}
\begin{center}
     \psfig{file=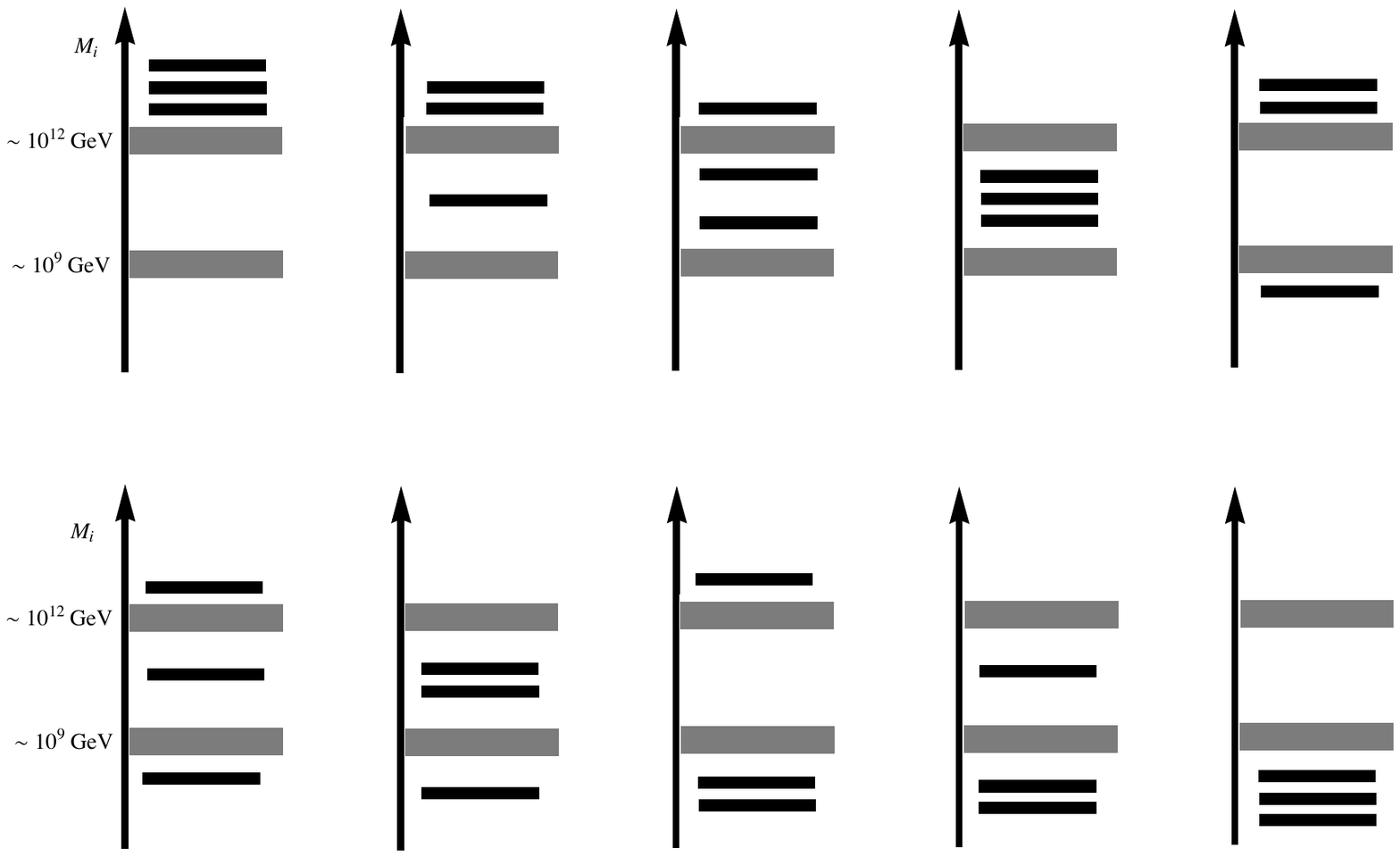,height=125mm,width=215mm}
     \vspace*{-53mm}
     \caption{The ten different three RH neutrino mass patterns requiring 10
                      different sets of Boltzmann equations for the calculation of the asymmetry \cite{problem}.}
\end{center}
\end{figure}
In addition, the requirement that all RH neutrino masses do not
fall in a transition regime becomes clearly  much more restrictive.

Moreover  new effects arise when heavy neutrino flavours are taken into account.
First, part of the asymmetry generated by a heavier RH neutrino species, 
the flavour orthogonal component,
escapes the washout from a lighter RH neutrino species \cite{bcst}.
Second, parts of the flavour asymmetries
(phantom terms) produced in the one or two flavour regimes
do not contribute to the total asymmetry at the production but
can contribute to the final asymmetry \cite{phantom}.

Therefore, it is necessary to extend the density matrix formalism  beyond the traditional
$N_1$-dominated scenario \cite{bcst,abada,riottodesimone} and account for heavy neutrino
flavours effects in order to calculate the final asymmetry for an arbitrary choice of the RH neutrino masses.
This is the main objective of this paper. At the same time we want to show how  Boltzmann equations  can be recovered
from the density matrix equations for the hierarchical RH neutrino mass patterns shown in Fig.~1 allowing
an explicit analytic calculation of the final asymmetry. In this way we will
confirm and extend results that were obtained within a simple quantum state collapse description.
For illustrative purposes, we will proceed in a
modular way, first discussing the specific effects in isolation
within simplified cases and then discussing the most general case that includes all effects.
The paper  is organised in the following way.

In Section 2 we discuss the derivation of the kinetic equations for the $N_1$-dominated scenario
in the absence of heavy neutrino flavours. This is useful both to show the
extension from  classical Boltzmann to density matrix equations and to highlight
some features that will prove to be quite important when, in the following Sections, we will include
heavy neutrino flavour effects.  In particular we show the existence of phantom terms and
 how the expression for the $C\!P$
asymmetry matrix can be unambiguously derived from the flavoured $C\!P$ asymmetries,
taking into account the different flavour compositions of the lepton and anti-lepton quantum
states produced in RH neutrino decays.

In Section 3  we start discussing the case where two heavy RH neutrino flavours
are involved directly in the generation of the asymmetry, considering
a simplified case with only two charged lepton flavours.
In this way we simplify the notation and we better highlight the main results.
In this  Section we are particularly interested to show two effects that specifically
arise when the interplay between heavy neutrino and charged lepton flavours is considered.
The first one is phantom leptogenesis \cite{phantom}. The second, that we call
projection effect, is how part of the asymmetry generated
by a heavy RH neutrino, the component orthogonal to the heavy neutrino flavour
associated to a lighter RH neutrino,
is not washed out by the inverse processes of the latter \cite{bcst,nir}.
We also show that these two effects, phantom leptogenesis and projection effect,
in general combine with each other.

In Section 4 we extend the discussion to the general case with three heavy neutrino
flavours and three charged lepton flavours. In this section we finally obtain
general density matrix equations for the calculation of the asymmetry
for an arbitrary choice of the RH neutrino masses.
From these equations we derive the classical Boltzmann equations for
a particularly interesting case: the two RH neutrino model. The derivation can be
easily extended to all ten hierarchical RH neutrino mass patterns shown in Fig.~1.
In Section 5 we draw the conclusions.

%%%%%%%%%%%%%%%%%%%%%%%%%%%%%%%%%%%%%%%%%%%%%%%%%%%%%%%%%%%%
\section{Kinetic equations for the $N_1$ dominated scenario}
%%%%%%%%%%%%%%%%%%%%%%%%%%%%%%%%%%%%%%%%%%%%%%%%%%%%%%%%%%%%

We discuss leptogenesis within a minimal type I seesaw mechanism with three RH neutrino species,
$N_1$, $N_2$ and $N_3$, with masses $M_1 \leq M_2 \leq M_3$ respectively where
one adds  right-handed neutrinos $N_{iR}$ to the SM lagrangian with Yukawa couplings $h$
and a Majorana mass term that violates lepton number
\be\label{lagrangian}
\mathcal{L}  =  \mathcal{L}_{\rm SM}
+i \, \overline{N_{iR}}\g_{\m}\partial^{\m} N_{iR} -
\overline{\ell_{\a L}}\,h_{\a i}\, N_{iR}\, \tilde{\F} -
 {1\over 2}\,M_i \, \overline{N_{iR}^c} \, N_{iR} + {\rm h.c.} \nonumber  \, \,\,  .
\ee
After spontaneous symmetry breaking, a neutrino Dirac mass term $m_D=v\,h$
is generated by the  Higgs vev $v$. In the seesaw limit, $M\gg m_D$, the spectrum
of neutrino masses splits into a light set given by the eigenvalues $m_1<m_2<m_3$
of the neutrino mass matrix
\be\label{seesaw}
m_{\nu} = - m_D\,{1\over M}\,m_D^T \, ,
\ee
and into a heavy set $M_1 < M_2 < M_3$ coinciding to a good
approximation with the eigenvalues of the Majorana mass matrix
corresponding to eigenstates $N_i\simeq N_{i R} + N_{i R}^c$.

In this section we review the main steps underlying the derivation of
the kinetic equations in leptogenesis when heavy neutrino flavours are neglected,
assuming that only the lightest RH neutrino decays and inverse processes
contribute to the final asymmetry: the traditional $N_1$-dominated scenario.

We first derive the Boltzmann (rate) equations and  then we extend them
writing the density matrix equations, accounting for quantum decoherence,
flavour oscillations and gauge interactions.
This discussion will prove to be useful not only to setup the notation
but also to highlight some basic
features of the kinetic equations in leptogenesis that will be relevant
when we will include heavy neutrino flavour effects in the next section.

We will neglect different effects, processes and corrections that
have been studied during the last years and that will not play a
relevant role in our discussion. These include for example
$\D L=2$ washout \cite{DL2, giudice},
$\D L=1$ scatterings \cite{early},
momentum dependence \cite{momentum},
thermal corrections \cite{giudice,thermal},
flavour coupling from the Higgs and quark asymmetries
\cite{bcst,spectators,nardi,N2dom,phantom},
quantum kinetic effects \cite{qke}.

We will moreover always assume vanishing pre-existing asymmetry though
notice that the results that we will obtain in Section 3 on the
projection effect,  are also important in order to describe the evolution
of a non-vanishing  pre-existing asymmetry \cite{nir,problem}.

\subsection{Boltzmann equations}

If we indicate with $\G_1$ the decay rate of the lightest RH neutrinos into leptons, $N_1\ra {\ell_1}+\Phi^{\dagger}$,
and with $\bar{\Gamma}_1$ the decay rate into anti-leptons,  $N_1\ra {\bar{\ell}_1}+\Phi$,
we can introduce the decay term $D_1$ and the washout term $W_1$ given respectively by
\be\label{DW}
D_1(z) \equiv {\G_1+\bar{\G}_1 \over H\,z}  = K_1\,z\,
\left\langle {1\over\gamma_1} \right\rangle  \;\;\;\;\; \mbox{\rm and} \;\;\;\;\;
W_1(z) \equiv {1\over 2}\,{\G_1^{ID}+\bar{\G}_1^{ID} \over H\,z} = {1\over 4}\,K_1 \,{\cal K}_1(z)\,z^3 \, ,
\ee
where $z\equiv M_1/T$, $K_1 \equiv (\G_1+\bar{\G}_1)_{T=0}/H_{T=M_1}$ is the decay parameter,
$H$ is the expansion rate
and the averaged dilution factor, in terms of the Bessel functions, is given by
$\left\langle {1/\gamma_1} \right\rangle = {{\cal K}_{1}(z) / {\cal K}_{2}(z)}$.
Under the fore-mentioned assumptions and approximations and considering
the unflavoured regime, the calculation of the asymmetry is described by
the most traditional set of kinetic equations for leptogenesis from the decays of
the lightest RH neutrinos $N_1$  \cite{early,bcst,cmb,giudice,pedestrians}
\begin{eqnarray}
{dN_{N_1}\over dz} & = & -D_1\,(N_{N_1}-N_{N_1}^{\rm eq}) \;,
\hspace{25mm}  \label{dlg1} \\
{dN_{B-L}\over dz} & = & \varepsilon_1\,D_1\,(N_{N_1}-N_{N_1}^{\rm eq})- W_1\,N_{B-L}  \, ,
\label{dlg2}
\end{eqnarray}
where with $N_X$ we indicate any particle number or asymmetry $X$
calculated in a portion of co-moving volume containing one heavy neutrino
in ultra-relativistic thermal equilibrium, in a way that $N^{\rm eq}_{N_i}(T\gg M_i)=1$.
In this way the baryon-to-photon number ratio at recombination is related to
the final $B-L$ asymmetry by
\be
\eta_B= a_{\rm sph}\,{N_{B-L}^{\rm f}\over N_{\gamma}^{\rm rec}}\simeq 0.01\,N_{B-L}^{\rm f} \, ,
\ee
to be compared with the  value measured from the CMB anisotropies observations \cite{WMAP7}
\be\label{etaBobs}
\eta_B^{\rm CMB} = (6.2 \pm 0.15)\times 10^{-10} \, .
\ee
Let us very shortly recall the basic steps for the derivation of the Eq.~(\ref{dlg2})
for the $B-L$ asymmetry. Ignoring the reprocessing action of sphalerons, we can write
\be\label{dNBmLdz}
{dN_{B-L}\over dz} = {dN_{\bar{\ell}_1}\over dz}-{dN_{\ell_1} \over dz} \,  .
\ee
 The net production rate of leptons and anti-leptons is then given by
the difference between the production rate due to decays and the
depletion rate due to inverse decays, for leptons
\be\label{dNl1}
{dN_{\ell_1} \over dz}= {\G_1\over H\, z}\,N_{N_1}-{\G_1^{ID}\over H\, z}\,N_{{\ell}_1}
\ee
and for anti-leptons
\be\label{bardNl1}
{dN_{\bar{\ell}_1} \over dz}= {\bar{\G}_1\over H\, z}\,N_{N_1}-{\bar{\G}_1^{ID}\over H\, z}\,N_{\bar{\ell}_1} \, .
\ee
The inverse decay rates are related to the decay rates by \cite{early}
\footnote{This expression directly accounts for the resonant $\D L=2$ contribution
that is needed not to violate the Sakharov condition on the necessity of a
departure from thermal equilibrium for the generation of an asymmetry \cite{DL2}.
Here we are interested in showing the separate Boltzmann equations for lepton anti-lepton numbers
that we will use in the next subsection to derive the $C\!P$ violating term
in the density matrix equations.}
\be
\G_1^{ID}=\G_1\,{N_{N_1}^{\rm eq} \over N_{\ell_1}^{\rm eq}} \hspace{5mm}
{\rm and} \hspace{5mm}
\bar{\G}_1^{ID}=\bar{\G}_1\,{N_{N_1}^{\rm eq} \over N_{\bar{\ell}_1}^{\rm eq}} \, ,
\ee
where $N_{\ell_1}^{\rm eq}=N_{\bar{\ell}_1}^{\rm eq} \equiv N_{\ell}^{\rm eq} = 1$
is the number of leptons ${\ell}_1$ and of anti-leptons  $\bar{\ell}_1$
in thermal equilibrium for vanishing asymmetry.
The number of leptons and anti-leptons can then be recast as
\bea\label{Nell1}
N_{\ell_1} & = & {1\over 2}\,\left({N_{\ell_1}+N_{\bar{\ell}_1}}\right)
+{1\over 2}\,\left({N_{\ell_1}-N_{\bar{\ell}_1}}\right) %\\ \nonumber
            =  N_{\ell}^{\rm eq}- {1\over 2}\,N_{B-L} + {\cal O}(N_{B-L}^2)
\eea
and
\bea\label{Nbarell1}
N_{\bar{\ell}_1} & = & {1\over 2}\,\left({N_{\ell_1}+N_{\bar{\ell}_1}}\right)-{1\over 2}\,\left({N_{\ell_1}-N_{\bar{\ell}_1}}\right) %\\ \nonumber
            =  N_{\ell}^{\rm eq} + {1\over 2}\,N_{B-L} + {\cal O}(N_{B-L}^2) \, .
\eea
Inserting these last expressions into the Eq.~(\ref{dNBmLdz}) and neglecting terms ${\cal O}(N_{B-L}^2)$,
the Eq.~(\ref{dlg2}) is obtained, with $D_1$ and $W_1$ given by the Eqs.~(\ref{DW}).

The solution for the final asymmetry has a very simple analytical expression \cite{pedestrians}
\be
N_{B-L}^{\rm f} = \ve_1\,\k(K_1) \, , \,\,  \hspace{5mm} \mbox{\rm with} \hspace{2mm}
\k(x) \equiv  \frac{2}{x \, z_{\rm B}(x)}\left[1-{\rm exp}\left(-\frac{1}{2}\,x \,
z_{\rm B}(x)\right)\right]\, ,
\ee
where $\k(K_1)$ is the final efficiency factor that here we have written, for simplicity, in the case of initial thermal $N_1$-abundance.
For $K_1\gtrsim 3$,  the strong wash-out regime favoured by neutrino
oscillation experiments, the asymmetry is
generated in quite a narrow interval of temperatures around $T_{B1}\equiv M_1/z_{B1}$,
where $z_{B1}\equiv z_{B}(K_1) ={\cal O}(1\div 10)$ \cite{pedestrians}.

The unflavoured assumption, underlying the Eqs.~(\ref{dlg1}) and (\ref{dlg2}), proves to describe the correct final asymmetry only
for masses $M_1\gtrsim 10^{13}\,{\rm GeV}$ \cite{riottodesimone,beneke}.  In this range of masses, during all the
relevant period of the asymmetry production, the lepton and anti-lepton quantum states produced from the
decays of the $N_1$, that we will indicate respectively simply with $|1\rangle$ and $|\bar{1}\rangle$,
can be treated, in flavour space, as pure states between their production at decay and their absorption
at a subsequent inverse decay.
They can be expressed as a linear combination of flavour eigenstates ($\a=e,\mu,\t$) ,
\be
|1\rangle =
\sum_{\a}\,{\cal C}_{1\a}\,| \a \rangle \, , \;\;\;\;
{\cal C}_{1\a} \equiv  \langle \a| 1 \rangle  \;\;\;\; \,
\hspace{5mm}
\mbox{\rm and}
\hspace{8mm}
|\bar{1}\rangle =
\sum_{\a}\,\bar{{\cal C}}_{\bar{1}\bar{\a}}\,|\bar{\a} \rangle \, , \;\;
\bar{{\cal C}}_{\bar{1}\bar{\a}} \equiv  \langle \bar{\a}|\bar{1} \rangle \, .
\ee
Notice that in general, even though in order to simplify the notation we are indicating
the final anti-leptons  produced by the $N_1$ decays with $\bar{\ell}_1$, they do not
coincide with the $C\!P$ conjugated of the final lepton states. This means that, introducing the
$C\!P$ conjugated states
\bea\label{1bar1}
C\!P  |\bar{1}\rangle & = & \bar{{\cal C}}_{1\tau}\,|\tau \rangle +
\bar{{\cal C}}_{1 \tau^{\bot}_1}\,|\tau^{\bot}_1\rangle \, , \;\mbox{\rm with} \,\,\,
\bar{{\cal C}}_{1 \a} =  \bar{\cal C}_{\bar{1}\bar{\alpha}}^{\star}  ,
\eea
in general one has $\bar{\cal C}_{1\a} \neq {\cal C}_{1\a}$ \cite{nardi}.

It will prove useful to introduce  the branching ratios  $p_{1\a}\equiv |{\cal C}_{1\a}|^2$
and $\bar{p}_{1\a}\equiv |\bar{\cal C}_{1\a}|^2$ giving respectively the probabilities
that a lepton ${\ell}_1$ or an anti-lepton $\bar{\ell}_1$ is found either in a flavour eigenstate $\a$
or $\bar{\a}$ in a flavour measurement process. It is also useful to recast the branching ratios as
\be\label{deltap1alpha}
p_{1\a}=  p^0_{1\a} +  {\d p}_{1\a}\, ,\hspace{5mm}
\bar{p}_{1\a}= p^0_{1\a} +  {\d \bar{p}}_{1\a} \, ,
\ee
where, in general, the tree level values $p^0_{1\a}\neq (p_{1\a}+\bar{p}_{1\a})/2$ and, therefore,
in general, $\delta p_{1\a} \neq - \delta \bar{p}_{1\a}$.
The deviations from the tree level values, $\d p_{1\a}=p_{1\a}-p^0_{1\a}$ and $\d\bar{p}_{1\a} = \bar{p}_{1\a}-p^0_{1\a}$,
originate from the $C\!P$ violating contributions due to the interference with loop diagrams
(see discussion in the Appendix).

If the charged lepton interactions are negligible during the period of
generation of the asymmetry (one-flavour regime) , for $z\simeq z_B$,
the lepton flavour compositions do not play any role since
the only other relevant interactions, the gauge interactions, are flavour blind and
lepton and anti-lepton quantum states propagate coherently between
production from decays and absorption from inverse decays. However, we will notice
that gauge interactions have some important interplay with lepton flavour compositions.
This situation is realised for $M_1\gtrsim 10^{13}\,{\rm GeV}$.
On the other hand, for masses $10^{12}\,{\rm GeV} \gg M_1 \gg 10^{9}\,{\rm GeV}$,
the coherent evolution of the $|1\rangle$ and $|\bar{1}\rangle$
quantum states breaks down before they can inverse
decay interacting with the Higgs bosons, due to collisions with right-handed tauons.
At the inverse decays, lepton quantum states can then be described as an incoherent mixture of
tauon eigenstates $|\tau\rangle$ and of  $|{\tau}^{\bot}_1\rangle$  quantum states.
These second ones are a coherent superposition of muon and electron eigenstates that can be regarded as the
projection of the lepton quantum states $|1\rangle$ on the plane orthogonal to the tauon flavour
(see Fig.~2).
\begin{figure}
\begin{center}
     \hspace*{10mm}
     \psfig{file=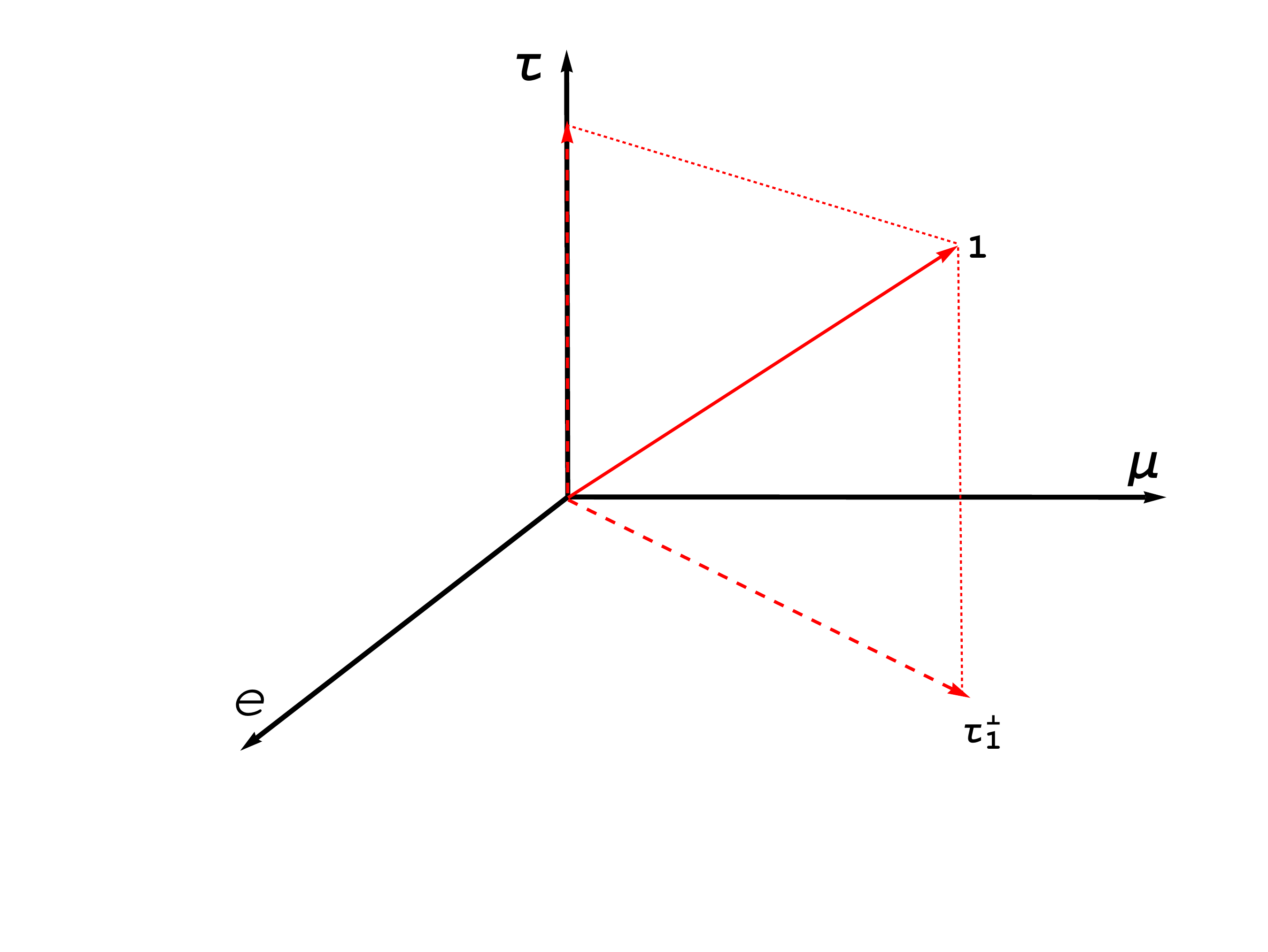,height=72mm,width=77mm}
     \vspace*{-8mm}
     \caption{For $10^{12}\,{\rm GeV} \gg M_1 \gg 10^{9}$GeV,
     the lepton quantum states $|1\rangle$ can be treated as an incoherent mixture of a $\tau$
                     and of a $\tau^{\bot}_1$ component
                     during the generation of the asymmetry and a two fully flavoured regime applies.}
\end{center}
\end{figure}
In this  two-fully flavored regime,  classical Boltzmann equations
can be still used as in the unflavored regime, with the difference, in general,
that now the flavour compositions of leptons and anti-leptons do play a role in the
generation of the asymmetry. In this case each single flavour asymmetry has to be
tracked independently and the total final $B-L$ asymmetry has to be calculated
after freeze-out as the sum of the two flavoured asymmetries,
a $\t$ asymmetry and a $\tau_1^{\bot}$ asymmetry.
To this extent, we have to introduce the flavoured $C\!P$ asymmetries
\be\label{epsial}
\ve_{i\a} \equiv
{\bar{p}_{i\a}\,\overline{\G}_{i} - p_{i\a}\,\G_{i}
\over \G_{i}+\overline{\G}_{i}} = {\bar{p}_{i\a}+p_{i\a}\over 2}\,\ve_{i}- {\Delta p_{i\a} \over 2}
\,  ,
\ee
where we defined $\Delta p_{i\a}\equiv p_{i\a}-\bar{p}_{i\a}$ and
all other quantities are a straightforward generalisation of
the quantities previously defined for the lightest RH neutrino species $N_1$
to the case of a generic RH neutrino species $N_i$.
Since sphaleron processes conserve the quantities $\D_{\a}\equiv B/3-L_{\a}$ ($\a=e,\m,\t $) \cite{bcst},
these are the convenient independent variables to be used in the set of Boltzmann equations that can
be written as
\bea\label{flke}
{dN_{N_1}\over dz} & = & -D_1\,(N_{N_1}-N_{N_1}^{\rm eq}) \, ,
\hspace{52mm}  \\ \nonumber
{dN_{\D_{\t}}\over dz} & = &
 \ve_{1\t}\,D_1\,(N_{N_1}-N_{N_1}^{\rm eq}) -\,p_{1\t}^{0}\,W_1\,N_{\D_{\t}} \, , \\  \nonumber
{dN_{\D_{\tau_1^{\bot}}}\over dz} & = &
\ve_{1{\tau}^{\bot}_1}\,D_1\,(N_{N_1}-N_{N_1}^{\rm eq})-p_{1{\tau}^{\bot}_1}^{0}\,W_1\,N_{\D_{{\tau}^{\bot}_1}} \, ,
\eea
where $p_{1{\tau}^{\bot}_1}^{0}\equiv p^0_{1e}+p^0_{1\mu}$
and $\ve_{1{\tau}^{\bot}_1}\equiv \ve_{1e}+\ve_{1\mu}$ and where we neglected terms ${\cal O}(\Delta p \, N_{\D_\a})$.

Using the decomposition of the flavoured $C\!P$ asymmetries in terms of $p^0_{i\a}$
and $\D p_{i\a}$ (cf. Eq.~(\ref{epsial})) and assuming strong washout for both flavours
(i.e. $K_{1\tau}, K_{1{\tau}^{\bot}_1} \gg 1$),
the final asymmetry is approximated by the expression
\be\label{twofully}
N^{\rm f}_{B-L} \simeq 2\,\ve_1\,\k(K_1) +
{\D p_{1\tau}\over 2}\,\left[\k(K_{1{\tau}^{\bot}_1})-\k(K_{1\tau})\right] \, ,
\ee
where $K_{i\a}\equiv p^0_{i\a}\,K_i$. This approximated expression
shows how, compared to the expression obtained in the unflavoured case, {\em large
lepton flavour effects can  arise only  when leptons and anti-leptons have a
different flavour composition}, for non-vanishing $\D p_{1\tau}$ 
\footnote{Notice that relaxing the assumption of strong washout for both flavours
one can only get an asymmetry  that is even closer to the unflavoured calculation.
Indeed in the limit of no washout  ($K_{1\tau}, K_{1{\tau}^{\bot}_1} \ll 1$)
one exactly recovers the unflavoured  expression for the final asymmetry.}.
In the Appendix we further discuss some interesting aspects and consequences
of this  point that is crucial for flavour effects to have a strong impact on the final asymmetry
and that, as we will see, will play a very important role in the results discussed in this paper.
The most extreme case is realized when $\ve_1=0$ \cite{nardi}.
In the unflavoured case this would imply a vanishing final asymmetry but
in the flavoured case it does not.  It should be indeed noticed that when
flavour effects are considered, $B-L$ violation is not a necessary condition for the generation
of a baryon asymmetry via leptogenesis,  it is sufficient to have a $\D_{\a}$ violation accompanied by
an asymmetric  washout between the two flavours, that in this context corresponds to
the requirement  of departure from  thermal equilibrium.

For $M_1\ll 10^9\,$ GeV muon interactions are able to break the coherent evolution
also of the $|{\tau}^{\bot}_1\rangle$  quantum states between decays and
inverse decays during the period of the generation of the asymmetry.
In this case a three-fully flavoured regime is realised
and the set of classical Boltzmann equations is a straightforward generalisation of that one
written in the two fully flavoured regime.  In the
$N_1$-dominated scenario with hierarchical RH neutrinos,
because of the lower bound $M_1\gtrsim 10^9\,$GeV
for successful leptogenesis \cite{di,cmb}, a three fully flavoured regime is not  relevant
for the calculation of the final asymmetry. On the other hand, in a $N_2$-dominated scenario,
a three flavoured regime has to be considered in the calculation of the lightest RH neutrino washout \cite{vives}.

\subsection{Density matrix equations}

Within a density matrix formalism \cite{bcst,abada,zeno,riottodesimone},
the description of leptogenesis is more general  than with classical Boltzmann equations,  since it makes possible to calculate the asymmetry in those intermediate regimes
where  lepton quantum states  interact with the thermal bath
via charged lepton interactions between decays and inverse decays
though not so efficiently that a quantum collapse approximation can be applied in a statistical description.
 In this case the ensemble of lepton quantum states cannot be described neither in terms of pure
states nor as an  incoherent mixture. Yukawa interactions and charged lepton interactions compete with each other in the determination of the average properties of the lepton quantum states. We will show that also gauge interactions  play
an active, though indirect, role.
A  statistical quantum-mechanical description of lepton flavour cannot treat leptons
as decoupled from the thermal bath.
Therefore, the concept of lepton quantum states itself is blurred since
one should consistently describe together leptons and thermal bath.  A density matrix formalism \cite{feynman}
is then particularly convenient since it  still allows to describe the leptonic subsystem in a separate way,
neglecting back-reaction effects and encoding the coupling with the thermal bath  in the evolution
of the off-diagonal terms  of the lepton density matrices.

Let us see how a density matrix equation for the $B-L$ asymmetry can be obtained
starting first from the case where charged lepton interactions are negligible.
In this case we just expect to reproduce the Eq.~(\ref{dlg2}).

Let us consider a simple two lepton flavour case able to describe the intermediate regime between the
unflavoured case and the two fully flavoured regime that are recovered as asymptotic limits.
The two relevant flavours  are then    $\tau$ and  $\tau_1^{\bot}$ (see Fig.~2).
In this two flavour space the flavour composition of the
lepton quantum states produced by the $N_1$ decays
can be written as $(\a=\tau,\tau^\bot_1)$
\bea\label{1bar1}
| 1\rangle & = &
{\cal C}_{1\tau}\,|\tau \rangle + {\cal C}_{1\tau^{\bot}_1}\,|\tau^{\bot}_1 \rangle \, , \;\;\;
{\cal C}_{1\a} \equiv  \langle \a| 1 \rangle  \, , \\
C\!P|\bar{1} \rangle & = & \bar{{\cal C}}_{1\tau}\,|\tau \rangle +
\bar{{\cal C}}_{1\tau^{\bot}_1}\,|\tau^{\bot}_1\rangle \, , \;\;\;
\bar{{\cal C}}_{1\a} \equiv  \langle \a | C\!P| \bar{1} \rangle \,  .
\eea
This definition can be straightforwardly generalised to the lepton quantum states
produced by a generic RH neutrino species $N_i$  that can be
written in terms of amplitudes ${\cal C}_{i\a}$ and $\bar{\cal C}_{i\a}$.
At tree level, the amplitudes ${\cal C}_{i\a}$ and $\bar{\cal C}_{i\a}$
are given by ($i=1,2,3$)
\be
 {\cal C}^0_{i\alpha} =   {h_{\a i} \over \sqrt{(h^{\dagger}\,h)_{ii}}}
 \hspace{6mm} \mbox{\rm and} \hspace{6mm}
 \bar{\cal C}^0_{i\alpha} = {h_{\a i} \over \sqrt{(h^{\dagger}\,h)_{ii}}}  \, .
 \ee
Including one-loop $C\!P$-violating corrections, these amplitudes become
\bea\label{Cialpha}
 {\cal C}_{i\alpha} & = & {1\over \sqrt{(h^{\dagger}\,h)_{ii}
  -2\,{\rm Re}(h^{\dagger}\,h\, \xi_u)_{ii}}}\left(h_{\a i}- (h \, \xi_u)_{\a i}\right) \, ,\\
 \bar{\cal C}_{i\alpha} & = & {1\over \sqrt{(h^{\dagger}\,h)_{ii}
 -2\,{\rm Re}(h^{\dagger}\,h\, \xi_v^{\star})_{ii}}}
 \left(h_{\a i} - (h\, \xi_v^{\star})_{\a i}\right)\, .\label{barCialpha}
\eea
This shows explicitly that the flavour compositions of leptons $\ell_i$ and of the
($C\!P$ conjugated) anti-leptons $\bar{\ell}_i$
are different, provided $\xi_v^{\star}\neq \xi_u$, which has to be the case for $C\!P$ violation to be
non-zero, as we show below.
We are following here  the notation and formalism introduced
in \cite{Buchmuller:1997yu} and more recently in \cite{Anisimov:2005hr}.
The one-loop corrections are included in the $\xi_u$ and $\xi_v$ functions, which are given by
\bea \label{uv}
\left[\xi_u(M_i^2)\right]_{ki} &\equiv & \left[ u (M_i^2) + M b (M_i^2) (h^{\dagger}\,h)^T M\right]_{ki} \, , \\ \nonumber
\left[\xi_v(M_i^2)\right]_{ki} &\equiv & \left[ v (M_i^2)+ M b (M_i^2)(h^{\dagger}\,h) M\right]_{ki} \, .
\eea
The first term on the right-hand side describes the self-energy correction, whereas the second
one is the vertex correction. Note that the mass matrix being diagonal, we simply have $M_{ki}=M_i\delta_{ki}$.
The $u$ and $v$ terms in Eq.~(\ref{uv}) are given by
\bea
u_{ki}(M_i^2) &=& \omega_{ki}(M_i^2)\left[ M_i \Sigma_{N,ki}(M_i^2) + M_k \Sigma_{N,ik}(M_i^2)\right] \, ,\\ \nonumber
v_{ki}(M_i^2) &=& \omega_{ki}(M_i^2)\left[ M_i \Sigma_{N,ik}(M_i^2) + M_k \Sigma_{N,ki}(M_i^2)\right] \, .
\eea
They depend on the propagator $\omega_{ik}$ and self-energy
$\Sigma_{N,ki}(M_i^2)=a(M_i^2)(h^{\dagger}\,h)_{ki}$, where $a$ is a loop
factor, both evaluated on mass-shell for the RH neutrino $N_i$.

It can be easily checked that
the difference of branching ratios, $\D p_{i\alpha} \equiv |{\cal C}_{i\alpha}|^2-|\bar{\cal C}_{i\alpha}|^2$,
does not vanish in general, implying different flavour compositions of leptons and anti-leptons.
This can be indeed expressed as
\bea\label{deltapexplicit}
|{\cal C}_{i\alpha}|^2-|\bar{\cal C}_{i\alpha}|^2 & &= {1\over (h^{\dagger}\,h)_{ii}}\sum_k\left\{4 M_i M_k\,
{\rm Im}\left[b_{ki}(M_i^2)\right] \,{\rm Im}\left[h^{\star}_{\alpha i} h_{\alpha k}(h^{\dagger}\,h)_{ik}\right]\right.
\\
&& +4 M_k \,{\rm Re}\left[\omega_{ki}(M_i^2)\right] \,
{\rm Im}\left[a(M_i^2)\right] \,{\rm Im}\left[h^{\star}_{\alpha i} h_{\alpha k}(h^{\dagger}\,h)_{ik}\right] \nonumber \\ \nonumber
&& +4 M_i \,{\rm Re}\left[\omega_{ki}(M_i^2)\right]\,
{\rm Im}\left[a(M_i^2)\right] \,{\rm Im}\left[h^{\star}_{\alpha i} h_{\alpha k}(h^{\dagger}\,h)_{ki}\right] \\ \nonumber
&&\left. -4 {|h_{\a i}|^2\over (h^{\dagger}h)_{ii}}M_k \left(M_i\,{\rm Im}\left[b_{ki}(M_i^2)\right]+
\,{\rm Re}\left[\omega_{ki}(M_i^2)\right]{\rm Im}\left[a(M_i^2)\right]\right)
\,{\rm Im}\left[(h^{\dagger}\,h)^2_{ik}\right] \right\}\nonumber  \, ,
\eea
where ${\rm Im}\left[a(M_i^2)\right] = -1/(16\pi)$, and
the imaginary part of the other loop factor $b(M_i^2)$ evaluated
on mass shell for the RH neutrino $N_i$ is given by
\be
{\rm Im}\left[b_{ki}(M_i^2)\right]={1\over 16 \pi M_i M_k} f(x_k/x_i) \, ,
\ee
where $x_i\equiv M^2_i/M^2_1$ and $f(x)=\sqrt{x}\left(1-(1+x)\log\left({1+x\over x}\right)\right)$.
Lastly, the real part of the propagator $\omega$, evaluated
on shell, is found to be
\be
{\rm Re}\left[\omega_{ki}(M_i^2)\right]={M_i(M_k^2-M_i^2) \over
(M_k^2-M_i^2)^2 +(M_k \Gamma_i- M_i \Gamma_k)^2} \, .
\ee
It can now be easily checked that the expression Eq.~(\ref{deltapexplicit})
consistently satisfies the decomposition  Eq.~(\ref{epsial}), explicitly
\be
|{\cal C}_{i\alpha}|^2-|\bar{\cal C}_{i\alpha}|^2 \equiv
\D p_{i\a}  = (p_{i\a}+\bar{p}_{i \a})\,\varepsilon_i - 2\,\varepsilon_{i\alpha}  \, ,
\ee
where $(p_{i\a}+\bar{p}_{i\a})  \simeq 2\, |h_{\a i}|^2/(h^{\dagger}\,h)_{ii}$,
the flavoured $C\!P$ asymmetries
\cite{cr}
\be
\ve_{i\a}= {3\over 16 \, \p(h^{\dagger}\, h)_{ii}}\sum_{j\neq i} \left\{ {\rm Im}\,
\left[ h_{\a i}^{\star}h_{\a j} (h^{\dagger}h)_{ij}\right] {\xi(x_j/x_i)\over \sqrt{x_j/x_i}}
+{2\over 3(x_j/x_i-1)}{\rm Im}\,\left[ h_{\a i}^{\star}h_{\a j} (h^{\dagger}h)_{ji}
\right]\right\} \, ,
\ee
with
\be\label{xi}
\xi(x)= {2\over 3}\,x\, \left[(1+x)\,\ln\left({1+x\over x}\right)-{2-x\over 1-x}\right]  \, ,
\ee
and, finally, the total $C\!P$ asymmetries
\be
\ve_i = {3\over 16 \, \p \, (h^{\dagger}\, h)_{ii}}\sum_{j\neq i} \, {\rm Im}\,
\left[ (h^{\dagger}h)^2_{ij}\right] {\xi(x_j/x_i)\over \sqrt{x_j/x_i}}  \, .
\ee
Let us now focus again on the $N_1$-dominated scenario. We can introduce
the quantum states  $|1^{\bot}\rangle$ and $C\!P |\bar{1}^{\bot}\rangle$ orthogonal,
in flavour space, respectively to the lepton quantum states
$|1 \rangle$ and $C\!P |\bar{1}\rangle$ and   with flavour compositions
\be
| 1^{\bot}\rangle  =
-{\cal C}^{\star}_{1\tau^{\bot}_1}\,|\tau \rangle +
{\cal C}^{\star}_{1\tau}\,|\tau^{\bot}_1 \rangle
\hspace{5mm} \mbox{\rm and} \hspace{5mm}
C\!P |\bar{1}^{\bot} \rangle  =  -\bar{\cal C}^{\star}_{1\tau^{\bot}_1}\,|\tau \rangle +
\bar{\cal C}^{\star}_{1\tau}\,|\tau^{\bot}_1\rangle  \,  .
\ee
 In the two flavour bases  $\ell_1$--$\ell_1^\bot$ and
 $C\!P(\bar{\ell}_1$--$\bar{\ell}_1^{\bot})$,
the lepton and anti-lepton density matrices
are respectively simply given by the projectors
$\rho^{\ell}_{ij}=\mathcal{P}_{ij}^{(1)}={\rm diag}(1,0)$ and
$\rho^{\bar{\ell}}_{i j}=\overline{\mathcal{P}}_{ij}^{(1)}={\rm diag}(1,0)$, where
$i,j=1,1^\bot$ if, for the time being, we assume that
there are no other leptons beyond the ${\ell}_1$'s produced by the RH neutrino decays
and that they are thermalised just by the Yukawa interactions. This is clearly not true either
if one starts from vanishing RH neutrino abundances or if Yukawa interactions are weak or
both. We will discuss in a moment how gauge interactions are able to thermalise the leptons and will
play a role, affecting the results on the asymmetries.
Notice that, for matrices,
we indicate the heavy neutrino flavour index with a superscript in round brackets.
Since we are dealing with $C\!P$ conjugated anti-lepton states,
we can use the same flavour indices for the matrix entries
of leptons and anti-leptons. However, it is important to notice that,
because of the different flavour  composition of leptons and anti-leptons,
the two bases do not coincide.

If we introduce the lepton and anti-lepton number density matrices, respectively
$N^{\ell}_{ij}\equiv N_{\ell_1} \, \rho^{\ell}_{i j}$ and
$N^{\bar{\ell}}_{ij}\equiv N_{\bar{\ell}_1} \, \rho^{\bar{\ell}}_{i j}$,
their evolution at $T\sim T_L$ is given by
 \be\label{dmke}
{dN^{\ell}_{ij}\over dz}  =
\left({\G_1\over H\,z}\,N_{N_1}- {\G_1^{ID}\over H\,z}\,N_{\ell_1} \right)\,\rho^{\ell}_{ij}  \, , \;\; \hspace{5mm}
{dN^{\bar{\ell}}_{ij}\over dz}  =
\left({\bar{\G}_1\over H\,z}\,N_{N_1}- {\bar{\G}_1^{ID}\over H\,z}\,N_{\bar{\ell}_1} \right)
\,\rho^{\bar{\ell}}_{ij} \, .
\ee
In order to obtain an equation for the total $B-L$ asymmetry matrix $N_{B-L}\equiv N^{\bar{\ell}}-N^{\ell}$,
we have first to write these two equations in the same flavour basis,
for convenience the lepton flavour basis $\t$--$\t_1^{\bot}$, and then subtract them.
The rotation matrices are then given by
\be
R_{\a i}^{(1)} = \left(\begin{array}{cc}
{\cal C}_{1\tau} & -{\cal C}^{\star}_{1\tau_1^{\bot}} \\
{\cal C}_{1\tau_1^{\bot}} &  {\cal C}^{\star}_{1\tau}
\end{array}\right)  \hspace{10mm}\mbox{\rm and} \hspace{10mm}
\bar{R}_{\a i}^{(1)} =
\left(\begin{array}{cc}
\bar{\cal C}_{1\tau} & -\bar{\cal C}^{\star}_{1\tau_1^{\bot}} \\
\bar{\cal C}_{1\tau_1^{\bot}} &  \bar{\cal C}^{\star}_{1\tau}
\end{array}\right)  \,  ,
\ee
for leptons and anti-leptons respectively.
Also notice that at tree level, corresponding to neglect $C\!P$ violation,
they  simply coincide, i.e.
\be\label{rotationmatrices}
R^{(1)0}_{\a i} = \left(\begin{array}{cc}
{\cal C}^0_{1\tau} & -{\cal C}^{0\star}_{1\tau_1^{\bot}} \\
{\cal C}^{0}_{1\tau_1^{\bot}} &  {\cal C}^{0\star}_{1\tau}
\end{array}\right) = \bar{R}^{(1)0}_{\a i} \, .
\ee
In the charged lepton flavour basis
one can finally write the equation for the $B-L$ asymmetry matrix as
\be\label{dNBmLab}
{dN^{B-L}_{\a\b} \over dz} =\bar{R}^{(1)}_{\a i}
\,{dN^{\bar{\ell}}_{i j}\over dz}\,
\bar{R}^{(1)\dagger}_{j\b} -
R^{(1)}_{\a i}\,{dN^{\ell}_{ij}\over dz} \, R^{(1)\dagger}_{j\b}  \, ,
\ee
whose trace gives the $B-L$ asymmetry $N_{B-L}$.
%Notice that
%we are now identifying the $\a$ and $\bar{\a}$ indexes since
%the lepton and anti-lepton flavor bases are $C\!P$ conjugated of each other.
In the charged lepton flavour basis the two projectors become
\be
\mathcal{P}_{\a\b}^{(1)}\equiv R^{(1) }_{\a i}
\,
\left(\begin{array}{cc}
1 & 0 \\
0 & 0
\end{array}\right)
\, R^{(1)\dagger}_{j\b}  =
\left(\begin{array}{cc}
p_{1\tau} & {\cal C}_{1\t}\,{\cal C}^{\star}_{1\tau_1^{\bot}} \\
{\cal C}^{\star}_{1\tau}\,{\cal C}_{1\tau_1^{\bot}} &  p_{1\tau_1^{\bot}}
\end{array}\right)   \, ,
\ee
\be
\overline{\mathcal{P}}^{(1)}_{\a\b}\equiv\bar{R}^{(1) }_{\a i}
\,
\left(\begin{array}{cc}
1 & 0 \\
0 & 0
\end{array}\right)
\, \bar{R}^{(1)\dagger}_{j\b}  =
\left(\begin{array}{cc}
\bar{p}_{1\tau} & \bar{\cal C}_{1\t} \,\bar{\cal C}^{\star}_{1\tau_{\bar{1}}^{\bot}}\\
\bar{\cal C}^{\star}_{1\tau}\,\bar{\cal C}_{1\tau_{\bar{1}}^{\bot}} &  \bar{p}_{1\tau_{\bar{1}}^{\bot}}
\end{array}\right)  \, ,
\ee
which, at tree level, simply coincide and are given by
\be\label{P0}
\mathcal{P}^{(1)0}_{\a\b}=
R^{(1)0}_{\a i}
\,
\left(\begin{array}{cc}
1 & 0 \\
0 & 0
\end{array}\right)
\, R^{(1)0\dagger}_{j\b} =
\left(\begin{array}{cc}
p^0_{1\tau} & {\cal C}^{0}_{1\t}\,{\cal C}^{0\star}_{1\tau_1^{\bot}} \\
{\cal C}^{0\star}_{1\tau}\,{\cal C}^{0}_{1\tau_1^{\bot}} &  p^0_{1\tau_1^{\bot}}
\end{array}\right)
={1\over (h^{\dagger}h)_{11}}
\left(\begin{array}{cc}
|h_{\t 1}|^2 & h_{\t 1}\, h_{\tau_1^{\bot} 1}^{\star} \\
h_{\t 1}^{\star}\, h_{\tau_1^{\bot} 1} & |h_{\tau_1^{\bot} 1}|^2
\end{array}\right) \, .
\ee
Using these results, we can now  rewrite Eq.~(\ref{dNBmLab}) as
\be
{dN^{B-L}_{\a\b} \over dz}  =
\left({\bar{\G}_1\over H\,z}\,N_{N_1}- {\bar{\G}_1^{ID}\over H\,z}\,N_{\bar{\ell}_1} \right)
\,\overline{\mathcal{P}}^{(1)}_{\a\b}-
\left({\G_1\over H\,z}\,N_{N_1}- {\G_1^{ID}\over H\,z}\,N_{\ell_1} \right)\,\mathcal{P}^{(1)}_{\a\b}   \, ,
\ee
that can be recast, using eqs.~(\ref{Nell1}) and (\ref{Nbarell1}) assuming thermal abundances, first as
\be
{dN^{B-L}_{\a\b} \over dz}  = \ve^{(1)}_{\a\b}\,D_1\,\left(N_{N_1}-N_{N_1}^{\rm eq}\right)
-W_1 \, N_{B-L}\,\left[ {\mathcal{P}^{(1)}_{\a\b} \,\G_1 +
\overline{\mathcal{P}}^{(1)}_{\a\b}\, \bar{\G}_1  \over \G_1 + \bar{\G}_1 }\right]
\ee
and then, neglecting terms ${\cal O}(\ve_1\,N_{B-L})$ and ${\cal O}(\Delta p\,N_{B-L})$, as
\be\label{NBmLabN1}
{dN^{B-L}_{\a\b} \over dz}   =
\ve^{(1)}_{\a\b}\,D_1\,(N_{N_1}-N_{N_1}^{\rm eq}) - W_1 \, N_{B-L}\, \mathcal{P}^{(1)0}_{\a\b} \,  .
\ee
Notice that this result has been obtained assuming that there are only ${\ell}_1$ leptons and $\bar{\ell}_1$ anti-leptons.
%The gauge interactions would also thermalise the ${\ell}_1^{\bot}$ and $\bar{\ell}_1^{\bot}$  but
%since they  are $C\!P$ conserving, there is no asymmetry produced in the flavour $1^{\bot}$ anyway.
%$\mathcal{P}^0$ was introduced in Eq.~(\ref{P0}), and we explicitly introduced the anti-commutator
%\{,\} structure in the equation because
%it is the relevant one in a density matrix equation.
Notice that we defined the $C\!P$ asymmetry matrix for the lightest  RH neutrino $N_1$ as
\be\label{asymmetrymatrix}
\ve^{(1)}= {\overline{\mathcal{P}}^{(1)}\,\overline{\G}_{1}- \mathcal{P}^{(1)}\,\G_{1}
\over \G_{1}+\overline{\G}_{1}} =
\ve_1\,{\overline{\mathcal{P}}^{(1)}+\mathcal{P}^{(1)} \over 2} -
{\D {\mathcal{P}}^{(1)} \over 2} \,  ,
\ee
where $\D {\mathcal{P}}^{(1)} \equiv  \mathcal{P}^{(1)}- \overline{\mathcal{P}}^{(1)}$.
This expression \cite{bcst} generalises the eq.~(\ref{epsial}) that is obtained
for the diagonal terms in the charged lepton flavour basis where
the diagonal terms simply correspond to the flavoured $C\!P$ asymmetries,
$\ve^{(1)}_{\a\a}=\ve_{1\a}$, while the off-diagonal terms  obey
$\ve^{(1)}_{\a\b}=(\ve^{(1)}_{\b\a})^{\star}$ and are not necessarily real.
This expression can be generalised to the $C\!P$ asymmetry matrix $\ve^{(i)}_{\a\b}$,
 of any RH neutrino species $N_i$ that
 in terms of the Yukawa couplings can be written as
\bea
\ve^{(i)}_{\a\b}&=& {3\over 32\p(h^{\dagger}h)_{ii}}\sum_{j\neq i} \left\{ {\rm i}\,
\left[ h_{\a i}h_{\b j}^{\star} (h^{\dagger}h)_{ji}
- h_{\b i}^{\star} h_{\a j} (h^{\dagger}h)_{ij}\right] {\xi(x_j/x_i)\over \sqrt{x_j/x_i}} \right. \nonumber\\
&&\left. \hspace{1cm}+{\rm i}\, {2\over 3(x_j/x_i-1)}\left[ h_{\a i}h_{\b j}^{\star} (h^{\dagger}h)_{ij}
- h_{\b i}^{\star} h_{\a j} (h^{\dagger}h)_{ji}\right] \right\} \, ,
\eea
where the $\xi$ function was defined in Eq.~(\ref{xi}).
This expression slightly differs from  that one in \cite{abada,riottodesimone}
(simply, there, the first term is minus the imaginary part of the first term
written here, so that the off-diagonal terms are real)
while it agrees with the expression given in \cite{beneke}.
%Notice that we have
%defined
%\be
%(W_1)_{\a\b} \equiv R^{0\dagger}_{\a i}
%\,
%\left(\begin{array}{cc}
%W_1 & 0 \\
%0 & 0
%\end{array}\right)
%\, R^0_{j\b}  = W_1\,\mathcal{P}^0_{\a\b}
%\ee
%so that
%\be\label{approximation}
%(\mathcal{P}^{0(1)} \, N^{B-L})_{\a\b} \simeq
%\left[R^{0(1)\dagger}
%\,
%\left(\begin{array}{cc}
%1 & 0 \\
%0 & 0
%\end{array}\right)
%\,
%\, R^{0(1)} \, R^{0(1)\dagger} \,
%\left(\begin{array}{cc}
%N_{B-L} & 0 \\
%0 & 0
%\end{array}\right)
%\, R^{0(1)} \right]_{\a\b}  =  N_{B-L}\,\mathcal{P}^{0(1)}_{\a\b}  \, .
%\ee
The  diagonal components of the Eq.~(\ref{dNBmLab}) can be explicitly written as
\bea\label{dNBmLee}
{dN^{B-L}_{\t\t}\over dz} & = &
\ve^{(1)}_{\t\t}\,D_1\,(N_{N_1}-N_{N_1}^{\rm eq})-
p_{1\t}^{0}\,W_1\,N_{B-L}  \, , \\ \label{dNBmLmm}
{dN^{B-L}_{\tau_1^{\bot}\tau_1^{\bot}}\over dz} & = &
\ve^{(1)}_{\tau_1^{\bot}\tau_1^{\bot}}\,D_1\,(N_{N_1}-N_{N_1}^{\rm eq})-
p_{1\tau_1^{\bot}}^{0}\,W_1\,N_{B-L}   \, .
\eea
Summing these two equations, one finally recovers the usual
Eq.~(\ref{dlg2}) for the total $B-L$ asymmetry $N_{B-L} = {\rm Tr}[N^{B-L}_{\a\b}]$,
%\be
%{dN_{B-L}\over dz}  =
%\ve_{1}\,D_1\,(N_{N_1}-N_{N_1}^{\rm eq})- W_1\,N_{B-L}  \, ,
%\ee
which is washed out in the usual way at the production. On the other hand, from
Eqs.~(\ref{dNBmLee}) and (\ref{dNBmLmm}), one finds the relation
\be
{1\over p^0_{1\t}}\,{dN^{B-L}_{\t\t}\over dz}-
{1\over p^0_{1\tau_1^{\bot}}}\,{dN^{B-L}_{\tau_1^{\bot}\tau_1^{\bot}}\over dz}
=-{\Delta p_{1\t}\over 2}\,\left({1\over p^0_{1\t}}+{1\over p^0_{1\tau_1^{\bot}}}\right)\,D_1\,(N_{N_1}-N_{N_1}^{\rm eq}) \,
\ee
which, together with Eq.~(\ref{dlg1}), forms a system of equations that can be solved analytically.
At low temperatures $T\ll T_{B1} = M_1/z_{B1} \ll M_1$, the final values are then found to be
\bea\label{NBmLttTB1}
N^{B-L,{\rm f}}_{\t\t}&\simeq& p^0_{1\t}\,N_{B-L}^{\rm f} - {\D p_{1\t}\over 2} \, N_{N_1}^{\rm in} \,  , \\ \nonumber
N^{B-L,{\rm f}}_{\tau_1^{\bot}\tau_1^{\bot}} &\simeq&  p^0_{1\tau_1^{\bot}}\,N_{B-L}^{\rm f} +
{\D p_{1\t}\over 2}\, N_{N_1}^{\rm in} \, .
\eea
This solution shows that the flavoured asymmetries contain  terms that escape the washout at
the production and are proportional to the initial abundance of RH neutrinos: these are the phantom terms \cite{phantom}.
If one only considers the one-flavour regime, where charged lepton interactions can be neglected
as we have done so far, the flavoured asymmetries are not themselves measured and the phantom terms
cannot give any physical effect, in particular they cannot affect the  baryon asymmetry.

We want now to consider the effect of charged lepton interactions and  of gauge interactions.
%$\mathcal{P}^{0(1)}$ and $N^{B-L}$ commute. % and therefore one has simply
%$\{\mathcal{P}^{0(1)},N^{B-L}\}_{\a\b}=2\,N_{B-L}\,\mathcal{P}^{0(1)}_{\a\b}$.
When charged
lepton interactions become effective, at $T \sim 10^{12}\,$GeV, tauon lepton
interactions start to be in equilibrium
breaking the coherence of the lepton quantum states.

Charged lepton interactions and gauge interactions are described by additional terms in Eqs.~(\ref{dmke}), which then
generalise into \cite{sigl,abada,thesis,beneke}
\bea\label{dmkewithclint}
{dN^{\ell}_{\a\b}\over dz}  & = &
{\G_1\over H\,z}\,N_{N_1} \, \mathcal{P}_{\a\b}^{(1)}
- {1\over 2}\,{\G_1^{ID}\over H\,z}\, \left\{\mathcal{P}^{(1)}  , N^{\ell} \right\}_{\a\b}
+\Lambda_{\a\b}  + G_{\a\b}  \,   , \\ \nonumber
{dN^{\bar{\ell}}_{\a\b}\over dz}  & = &
{\bar{\G}_1\over H\,z}\,N_{N_1} \, \overline{\mathcal{P}}^{(1)}_{\a\b}
- {1\over 2}\,{\bar{\G}_1^{ID}\over H\,z}\,
\left\{\overline{\mathcal{P}}^{(1)}  , N^{\bar{\ell}} \right\}_{\a\b}  + \bar{\Lambda}_{\a \b}  + \bar{G}_{\a \b}  \, .
\eea
where $\L_{\a\b}$ and $\bar{\L}_{\a\b}$ are the terms describing the effect of charged lepton interactions,
\bea
\Lambda_{\a\b}            & = &
- {\rm i}\,{{\rm Re}(\Lambda_{\t})\over H\, z}\left[\left(\begin{array}{cc}
1 & 0 \\
0 & 0
\end{array}\right),N^{\ell}\right]_{\a\b}
-{{\rm Im}(\L_{\t})\over H\, z}\,\left(\begin{array}{cc}
0 & N^{\ell}_{\tau\tau_1^{\bot}} \\
N^{\ell}_{\tau_1^{\bot}\tau} & 0
\end{array}\right) \,    , \\
\bar{\Lambda}_{\a \b}  & = &
 + {\rm i}\,{{\rm Re}(\Lambda_{\t})\over H\, z}\left[\left(\begin{array}{cc}
1 & 0 \\
0 & 0
\end{array}\right),N^{\bar{\ell}}\right]_{\a\b}
-{{\rm Im}(\L_{\t})\over H\, z}\,\left(\begin{array}{cc}
0 & N^{\bar{\ell}}_{\tau\tau_1^{\bot}} \\
N^{\bar{\ell}}_{\tau_1^{\bot}\tau} & 0
\end{array}\right) \, .
\eea

The real and imaginary parts of the tau-lepton self-energy are respectively given by \cite{weldon, cline}
\be
{\rm Re}(\L_{\t}) \simeq {f_{\t}^2\over 64} \, T \,  \hspace{10mm} \mbox{\rm and}
\hspace{10mm} {\rm Im}(\L_{\t}) \simeq 8\times 10^{-3} \, f_{\t}^2\, T \, ,
\ee
where $f_{\t}$ is the tauon Yukawa coupling.
The commutator structure in the third term on the RHS of Eq.~(\ref{dmkewithclint})
accounts for oscillations in flavor space driven by the real part of
the self energy, and the second terms damp of the off-diagonal terms
driven by the imaginary part of the self energy.
%These effects can also be conveniently expressed in terms of the Pauli matrices
%$\sigma_i,~i=1,2,3$, so that the evolution equation reduces to
%\bea
%{dN^{B-L}_{\a\b} \over dz}  & = &
%\ve^{(1)}_{\a\b}\,D_1\,(N_{N_1}-N_{N_1}^{\rm eq})-{1\over 2}\,W_1\,\left\{{\cal P}^{0(1)}, N^{B-L}\right\}_{\a\b} \\ %\nonumber
%& & -\,{{\rm Re}(\Lambda_{\t})\over H\, z} (\sigma_2)_{\a\b} N^{\ell +\bar{\ell}}_{\a\b}
%- {{\rm Im}(\L_{\t})\over H\, z} (\sigma_1)_{\a\b}\,N^{B-L}_{\a\b} \ ,
%\eea
The terms $G_{\alpha\beta}$ and $\bar{G}_{\alpha\beta} $
describe the gauge interactions and have the effect to thermalise
leptons and anti-leptons so that kinetic and chemical equilibrium
can be assumed during all the transition from the unflavoured regime to the
two fully flavoured regime. Since they are $C\!P$ conserving, they cannot
change the total and flavour asymmetries while thermalising the asymmetries but,
as we are going to discuss, they play an active, though indirect,
role in the final values of the asymmetries.

Let us  show  how, from the set of density matrix equations (\ref{dmkewithclint}), one can
derive correctly  both the one-flavour
 (cf. eq.~(\ref{NBmLabN1}))  and the (two) fully flavoured regime (cf. eq.~(\ref{flke})).

 In {\em the one-flavour case} we have seen that neglecting gauge interactions corresponds to have
$N^{\ell}= N_{{\ell}_1}\,\mathcal{P}^{(1)}$   and
$N^{\bar{\ell}}=N_{\bar{\ell}_1}\,\overline{\mathcal{P}}^{(1)}$, where
we had to assume that $N_{{\ell}_1}$ and $N_{\bar{\ell}_1}$
are thermalised by the same Yukawa  interactions, an assumption that
does not describe the case either when Yukawa interactions are weak or
if one starts from a non-thermal RH neutrino abundance.

If we now take into account the effect of  gauge interactions, these will thermalise not only the abundances
of the leptons ${\ell}_{1}$ and of the anti-leptons $\bar{\ell}_{1}$, independently of the strength of the Yukawa
interactions and of the RH neutrino abundance, but also the abundances of their orthogonal states
${\ell}_{1^{\bot}}$ and $\bar{\ell}_{1^{\bot}}$. Since they are flavour blind and $C\!P$ conserving,
their presence is described by an additional unflavoured term in the
lepton and anti-lepton abundance matrices that in this way get generalised as
\bea\label{nice1}
N^{\ell}  & =  & N_{\ell}^{\rm eq} \, I  + N_{\ell_1}\, \mathcal{P}^{(1)}
-{N_{\ell_1}+N_{\bar{\ell}_1}\over 2}\, {\mathcal{P}}^{(1)0} \,  , \\  \nonumber
N^{\bar{\ell}} & =  & N_{\ell}^{\rm eq} \, I  + N_{\bar{\ell}_1} \, \overline{\mathcal{P}}^{(1)}
 -  {N_{\ell_1}+N_{\bar{\ell}_1}\over 2} \, {\mathcal{P}}^{(1)0}   \ .
\eea
The third terms in the right-hand side describe how annihilations mediated by gauge interactions drag out of the
${\ell}_1$ and $\bar{\ell}_1$ their tree-level components, $C\!P$ conjugated of each other,
that are thermalised.
In this way the gauge interactions annihilations act as a sort of detector of the
differences of flavour compositions of leptons and anti-leptons, though they cannot measure
the flavour compositions themselves, as implied by the term $N_{\ell}^{\rm eq}\,I$ that is invariant under
rotations in flavour space.  
If we linearise $N_{{\ell}_1}$ and $N_{\bar{\ell}_1}$ using the eqs.~(\ref{Nell1})
and (\ref{Nbarell1}) respectively, they can be recast as
\footnote{Notice that now these equations also describe consistently the case of vanishing initial
RH neutrino abundance that would yield seemingly unphysical negative values of $N_{\ell_1}+N_{\bar{\ell}_1}$. Indeed
now negative values correspond to the production of orthogonal states ${\ell}_{1^{\bot}}$ and $\bar{\ell}_{1^{\bot}}$, considering that
$\mathcal{P}^{(1)}= I - \mathcal{P}^{(1)}_{\bot}$, $\mathcal{P}^{(1)0}= I - \mathcal{P}^{(1)0}_{\bot}$
and analogously for the anti-leptons.}
\footnote{
One could wonder whether instead of terms proportional to the tree level components 
$\mathcal{P}^{(1)0}$,
one should subtract  in the eqs.~(\ref{nice1}) terms proportional  
to the average components $ \left(\overline{\mathcal{P}}^{(1)} +
{\mathcal{P}}^{(1)} \right)/2$.  However, one can verify that
one would anyway obtain in the end the same result eq.~(\ref{unflavoured})  
unless ${\cal O}(\Delta P^2)$ terms.
Notice also that with this modification (and neglecting terms 
${\cal O}(N_{B-L}\,\D {\mathcal{P}}^{(1)})$) the expressions (\ref{nice})
can be written as
\bea\label{verynice}
N^{\ell}  & =  & N_{\ell}^{\rm eq} \, I  + \left({N_{\ell_1}+N_{\bar{\ell}_1}\over 2}  \right)\,
{\D\mathcal{P}^{(1)}\over 2}- {1\over 2}\, N_{B-L}\,
{\overline{\mathcal{P}}^{(1)} + {\mathcal{P}}^{(1)}  \over 2}    \,  ,  \\  \nonumber
N^{\bar{\ell}} & =  & N_{\ell}^{\rm eq} \, I -
\left({N_{\ell_1}+N_{\bar{\ell}_1}\over 2}  \right)\, {\D{\mathcal{P}}^{(1)}\over 2} 
+ {1\over 2}\, N_{B-L} \,   
{\overline{\mathcal{P}}^{(1)} + {\mathcal{P}}^{(1)}  \over 2}   \ ,
\eea
 respecting the constraint \cite{beneke} 
\be\label{constraint}
N^{\ell} - N^{\rm eq}_{\ell}\, I = 
-  (N^{\bar{\ell}} - N^{\rm eq}_{\ell} \, I) \,   ,
\ee
that would be  a matrix generalisation of the thermal equilibrium conditions Eqs. (12) and (13).  
}
\bea\label{nice}
N^{\ell}  & =  & N_{\ell}^{\rm eq} \, I  + \left({N_{\ell_1}+N_{\bar{\ell}_1}\over 2}  \right)\,
\d\mathcal{P}^{(1)}- {1\over 2}\, N_{B-L}\,\mathcal{P}^{(1)}   \,  ,  \\  \nonumber
N^{\bar{\ell}} & =  & N_{\ell}^{\rm eq} \, I +
\left({N_{\ell_1}+N_{\bar{\ell}_1}\over 2}  \right)\, \d\overline{\mathcal{P}}^{(1)} 
+ {1\over 2}\, N_{B-L} \,\overline{\mathcal{P}}^{(1)}      \ ,
\eea
where  we defined $\d {\mathcal{P}}^{(1)} \equiv {\mathcal{P}}^{(1)} - {\mathcal{P}}^{(1)0} $
and $\d\overline{\mathcal{P}}^{(1)} \equiv \overline{\mathcal{P}}^{(1)} - {\mathcal{P}}^{(1)0} $.
 From these equations one can find  an expression for the asymmetry matrix,
\be\label{difference}
N^{B-L} =
N_{B-L}\, {\mathcal{P}^{(1)}+\overline{\mathcal{P}}^{(1)} \over 2}
- \left({N_{\ell_1}+N_{\bar{\ell}_1}}\right) \,{\D{\mathcal{P}}^{(1)} \over 2}  \, ,
\ee
that has to be compared with the eq.~(\ref{asymmetrymatrix}) for the $C\!P$
asymmetry matrix: the first term is the usual contribution proportional to the total asymmetry,
while the second term is the contribution to the flavour asymmetry matrix coming
from the difference in flavour compositions yielding the phantom terms. Notice that the quantity
$(N_{\ell_1}+N_{\bar{\ell}_1})/2$ has to be regarded as a dynamical quantity, like
the total asymmetry $N_{B-L}$.
We can also write an expression for the sum
\be\label{sum}
N^{{\ell}+\bar{\ell}} \equiv N^{\ell} + N^{\bar{\ell}} = 2\,N^{\ell}_{\rm eq} \, I +
{N_{\ell_1}+N_{\bar{\ell}_1}\over 2} \,\left(\d\mathcal{P}^{(1)} +
\d\overline{\mathcal{P}}^{(1)} \right) -{N_{B-L}\over 2}\,\D\mathcal{P}^{(1)}  \,  .
\ee
Considering that in the tree level basis one has ($i_0, j_0 = 1_0, 1_0^{\bot}$)
\bea
 \d {\mathcal{P}}^{(1)}_{i_0 j_0} =
\left(\begin{array}{cc}
0 & \d p^{\star} \\
\d p & 0
\end{array}\right)  \,  ,
\eea
with $\d p = -{\cal C}^0_{1\tau_1^{\bot}}\,\d {\cal C}_{1\tau} + \,{\cal C}^0_{1\tau}\,\d {\cal C}_{1\tau_1^{\bot}} $
and $\d{\cal C}_{1\a}\equiv {\cal C}_{1\a}-{\cal C}^0_{1\a}$, one obtains the equalities
\be\label{equalities}
\left\{ \mathcal{P}^{(1)} , \d \mathcal{P}^{(1)} \right\} = \d \mathcal{P}^{(1)} + {\cal O}(\d \mathcal{P}^2)\, ,
 \hspace{5mm}
\left\{ \overline{\mathcal{P}}^{(1)} , \d \overline{\mathcal{P}}^{(1)} \right\} =
\d \overline{\mathcal{P}}^{(1)} +{\cal O}(\d \overline{\mathcal{P}}^2)\,  ,
\ee
and neglecting terms ${\cal O}(\ve\,\D P)$ and ${\cal O}(\d \mathcal{P}^2)$,
one arrives at the following equation
\be\label{unflavoured}
{dN^{B-L} \over dz}   =
\ve^{(1)} \,D_1\,(N_{N_1}-N_{N_1}^{\rm eq}) -
{1\over 2} \, W_1 \, {N_{\ell_1}+N_{\bar{\ell}_1}\over 2} \,\left(\overline{\mathcal{P}}^{(1)}-
{\mathcal{P}}^{(1)} \right) - W_1\, {N_{B-L}\over 2}\,
\left(\mathcal{P}^{(1)}+\overline{\mathcal{P}}^{(1)} \right)  \,  .
\ee
Using the eqs.~(\ref{equalities}), it can be also recast more compactly as
\footnote{We wish to thank M.~Herranen and B.~Garbrecht
for pointing out to us that the eq.~(\ref{unflavoured2}) implies some wash-out of the phantom
terms and that, therefore, is not equivalent to the eq.~(\ref{NBmLabN1}) when the differences between lepton and anti-lepton flavour compositions are taken into account.}
\be\label{unflavoured2}
{dN^{B-L} \over dz}   =
\ve^{(1)}\,D_1\,(N_{N_1}-N_{N_1}^{\rm eq}) -
{1\over 2} \, W_1 \, \left\{{\mathcal{P}}^{(1)0}, N^{B-L}\right\}  .
\ee
The Eq.~ (\ref{unflavoured}) implies that,
having accounted for the unflavoured thermal bath from gauge interactions,
phantom terms are washed out contrarily to the previous calculation where it was neglected.
However, non trivially,  the  wash-out term acting on phantom terms
is half compared to that one acting on the total asymmetry.  Let us show this result explicitly,
finding the solutions for the diagonal components in the charged lepton flavour basis,
$N^{B-L}_{\t\t}$ and  $N^{B-L}_{\t_1^{\bot}\t_1^{\bot}}$. If one first considers the
eq.~(\ref{unflavoured2}) in the tree level basis, in this basis the decomposition of
$\ve^{(1)}$ in the right-hand side of eq.~(\ref{asymmetrymatrix}) specialises into
\be
\ve^{(1)}_{i_0 j_0} =
\left(\begin{array}{cc}
\ve_1 & 0 \\
0 & 0
\end{array}\right)   +
\left(\begin{array}{cc}
0 & \D\ve^{\star} \\
\D\ve & 0
\end{array}\right) \, ,
\ee
where $\D\ve=(\d\bar{p}-\d p)/2 \equiv -\D p/2$. In this way in this basis the $1_0 1_0$ 
term is just the
total asymmetry $N_{B-L}$ that gets washed out by $W_1$. Instead the off-diagonal terms,
upon rotation to the charged lepton flavour basis, give the phantom terms that are
washed by $W_1/2$. In this way, in the charged lepton flavour basis, one finds
\bea\label{NBmLttTB2}
N^{B-L,{\rm f}}_{\t\t}&\simeq& p^0_{1\t}\,N_{B-L}^{\rm f}
- {\D p_{1\t}\over 2} \, \k(K_1/2) \,  , \\ \nonumber
N^{B-L,{\rm f}}_{\tau_1^{\bot}\tau_1^{\bot}} &\simeq&  p^0_{1\tau_1^{\bot}}\,N_{B-L}^{\rm f} 
+ {\D p_{1\t}\over 2}\, \k(K_1/2) \, .
\eea
This result  confirms the presence of phantom terms but it
also clearly shows how the effect of the gauge interactions annihilations in detecting the
differences between lepton and anti-lepton flavour compositions results into a wash-out of the
phantom terms, though with a wash-out rate that is halved compared to the wash-out rate
acting on the total asymmetry.

Let us now consider the (two) fully flavoured regime. This can be recovered more conveniently considering that
in the eqs.~(\ref{dmkewithclint})  the off-diagonal terms are damped
by the charged lepton interactions \cite{bcst}. Therefore, one has that $N^{\ell}$ and $N^{\bar{\ell}}$ are diagonal
in the charged lepton flavour basis, so that $N^{\ell}_{\a\b} ={\rm diag}(N^{\ell}_{\t\t},N^{\ell}_{\t_1^{\bot}\t_1^{\bot}})$
and $N^{\bar{\ell}}_{\a\b} ={\rm diag}(N^{\bar{\ell}}_{\t\t},N^{\bar{\ell}}_{\t_1^{\bot}\t_1^{\bot}})$.
The gauge interactions thermalise the $\tau$ and the $\tau_1^{\bot}$ abundances.
In this way, taking the diagonal
components, one straightforwardly recovers the eqs.~(\ref{flke}).

Notice that one could also try to get this result  from a closed differential equation for $N^{B-L}_{\a\b}$.
Subtracting the two equations (\ref{dmkewithclint}) one obtains
\bea\label{NB-Lab}
{dN^{B-L}_{\a\b} \over dz}  & = &
\ve^{(1)}_{\a\b}\,D_1\,N_{N_1}
- {1\over 2}\, D_1 \, \left[ {\overline{\G}_1^{\rm ID} \over \G_1 + \overline{\G}_1 }\left\{\overline{\mathcal{P}}^{(1)},N^{\bar{\ell}}\right\}_{\a\b} - {\G_1^{\rm ID} \over \G_1 + \overline{\G}_1 }\,\left\{ \mathcal{P}^{(1)} , N^{\ell} \right\}_{\a\b}   \right]   \\ \nonumber
& & + \Delta \Lambda_{\a\b} + \D G_{\a\b} \,  .
\eea
Recasting then
\be
N^{\ell}= {N^{\ell}+N^{\bar{\ell}}\over 2} - {N^{B-L}\over 2} \, \hspace{5mm} {\rm and}
\hspace{5mm}
N^{\bar{\ell}}= {N^{\ell}+N^{\bar{\ell}}\over 2} + {N^{B-L}\over 2}  \, ,
\ee
one obtains first
\bea\label{NB-Lab3}
{dN^{B-L}_{\a\b} \over dz}  & = &
\ve^{(1)}_{\a\b}\,D_1\,N_{N_1} - {1\over 4}\,D_1\,{N_{N_1}^{\rm eq}\over N_{\ell}^{\rm eq}}\,\left\{ \ve^{(1)}_{\a\b} , {N^{\ell + \bar{\ell}}} \right\}_{\a\b}   \\ \nonumber
& & - {1\over 4}\, D_1 \, \left[ {\overline{\G}_1^{\rm ID} \over \G_1 + \overline{\G}_1 }\left\{\overline{\mathcal{P}}^{(1)},N^{B-L}\right\}_{\a\b} + {\G_1^{\rm ID} \over \G_1 + \overline{\G}_1 }\,\left\{ \mathcal{P}^{(1)} , N^{B-L} \right\}_{\a\b}   \right]
\\  \nonumber
& & + \Delta \Lambda_{\a\b} + \D G_{\a\b} \,  .
\eea
and then, neglecting terms ${\cal O}(\D P \, N_{B-L})$,
\bea\label{NB-Lab2}
{dN^{B-L}_{\a\b} \over dz}  & = &
\ve^{(1)}_{\a\b}\,D_1\,N_{N_1} - {1\over 4}\,D_1\,{N_{N_1}^{\rm eq}\over N_{\ell}^{\rm eq}}\,\left\{ \ve^{(1)}_{\a\b} , {N^{\ell + \bar{\ell}}} \right\}_{\a\b}   
- {1\over 2}\, W_1 \, \left\{{\mathcal{P}}^{0(1)}, N^{B-L}\right\}_{\a\b}
\\ & & +
 {\rm i}\,{{\rm Re}(\Lambda_{\t})\over H\, z}\left[\left(\begin{array}{cc}
1 & 0 \\
0 & 0
\end{array}\right),N^{\ell +\bar{\ell}}\right]_{\a\b}
-{{\rm Im}(\L_{\t})\over H\, z}\,\left(\begin{array}{cc}
0 & N^{B-L}_{\tau\tau_1^{\bot}} \\ \nonumber
N^{B-L}_{\tau_1^{\bot}\tau} & 0
\end{array}\right)  + \Delta G_{\a\b}   \,  .
\eea
 A Boltzmann equation for the quantity $N_{\ell +\bar{\ell}}$ is given by
\be
{dN^{\ell +\bar{\ell}}_{\a\b} \over dz} \simeq
-{{\rm Re}(\Lambda_{\t})\over H\, z}(\sigma_2)_{\a\b} N^{B-L}_{\a\b} -S_g \,
(N^{\ell +\bar{\ell}}_{\a\b}-2\,N_{\ell}^{\rm eq }\delta_{\a\b}) \, ,
\ee
where $S_g\equiv \Gamma_g/(Hz)$ accounts for gauge interactions. As shown in \cite{beneke}, this term has the effect of damping the flavour oscillations. This is because the gauge
interactions force $N^{\ell +\bar{\ell}}_{\a\b} \simeq 2\,N_{\ell_1}^{\rm eq }\delta_{\a\b}$ \cite{bcst,thesis,beneke}, as it can be seen explicitly from eq.~(\ref{sum}). This in turn 
makes in a way that the oscillatory term becomes negligible and that
the second term on the right-hand side can be approximated with the usual inverse decay 
$C\!P$ violating term, obtaining in the end
 \be\label{fullyflavoured}
{dN^{B-L}_{\a\b} \over dz}   =
\ve^{(1)}_{\a\b}\,D_1\,(N_{N_1}-N_{N_1}^{\rm eq})-{1\over 2}\,W_1\,\left\{{\cal P}^{0(1)}, N^{B-L}\right\}_{\a\b}
- {{\rm Im}(\L_{\t})\over H\, z} (\sigma_1)_{\a\b}\,N^{B-L}_{\a\b} \  ,
\ee
that generalises the Eq.~(\ref{unflavoured2}).
When the off-diagonal terms are fully damped,
one again correctly obtains the  Eqs.~(\ref{flke}) in the fully flavoured regime
and the usual Eq.~(\ref{dlg2}) for the total asymmetry  in the unflavoured regime.
Notice that we have now shown that the eq.~(\ref{fullyflavoured}) holds 
also starting from the Eqs.~(\ref{nice1}), taking into account differences between lepton and anti-lepton flavour compositions. This is because they anyway respect 
the approximation $N^{\ell +\bar{\ell}}_{\a\b} \simeq 2\,N_{\ell_1}^{\rm eq }\delta_{\a\b}$ 
(cf. \ref{sum}). 
 
 Suppose now that the asymmetry was generated in the unflavoured regime at temperatures $T\gg 10^{12}\,{\rm GeV}$.
 Let us indicate with $T_{\star}\ll 10^{12}\,{\rm GeV}$ that value of the temperature below which
one can approximate, with the desired precision,
the lepton quantum states as a fully incoherent mixture of $|\tau\rangle$
and $|\tau^{\bot}_1 \rangle$ quantum states corresponding to a complete damping
of the off-diagonal terms in the lepton density matrix (analogously for anti-leptons). This means that
 for $T \lesssim T_{\star}$  the $\t$  and  $\t_1^{\bot}$ lepton asymmetries,
 given by the diagonal entries of $N^{B-L}_{\a\b}$,
 are fully measured by the  thermal bath and reprocessed by sphaleron processes conserving
 the $\D_{\t}$ and $\D_{\tau_1^{\bot}}$ asymmetries, so that at $T_{\star}$ one has
 $N_{\D_{\t}}^{T_{\star}}=N^{B-L}_{\t\t}(z_B)$ and $N_{\D_{\tau_1^{\bot}}}^{T_{\star}}=N^{B-L}_{\tau_1^{\bot}\tau_1^{\bot}}(z_B)$.
 However, notice that since in the case $M_1\gg 10^{12}\,$GeV the total $B-L$ asymmetry got already produced and
frozen in the unflavoured regime, this fully flavoured regime stage does not affect the final total $B-L$ asymmetry.
Therefore, phantom terms do not contribute to the final asymmetry because
they cancel with each other. In other words, within the $N_1$-dominated scenario, phantom terms have
no effect on the final asymmetry
\footnote{On the other hand, as discussed, if one considers $10^9\,{\rm GeV}\ll M_1\ll 10^{12}\,$GeV, the two fully flavoured regime holds during the period of leptogenesis and the density matrix equations
reduce to the set of classical Boltzmann equations Eqs.~(\ref{flke}).
The terms in the flavoured asymmetries coming from $C\!P$ violating terms due to a different
flavour composition of leptons and anti-leptons are still present but they are not phantom, since they
are measured directly at production and undergo washout. Therefore, if there is
a flavour-asymmetric production, they  contribute to the final asymmetry, yielding
the second term in the Eq.~(\ref{twofully}), and can even dominate.}.
Therefore, phantom terms  do not have any consequence on the
final asymmetry in the $N_1$-dominated scenario,
in the absence of heavy neutrino flavour effects.

Our results show explicitly  the presence of the phantom terms extending 
previous results where this had not been noticed \cite{riottodesimone,beneke}. 
In particular, in \cite{riottodesimone}, the lepton and anti-lepton density matrices were 
assumed to be diagonalisable  in  bases that are $C\!P$ conjugated of each other
precluding the derivation of the phantom terms.
%switching to the
%charged lepton flavour basis from the two heavy neutrino flavour
%bases \cite{beneke}. In \cite{riottodesimone} the equations were written directly
%in the charged lepton flavour basis and it was shown that the usual solution for the
%total asymmetry is recovered while the solutions
 %for the flavoured asymmetries   were not explicitly
%written and the phantom terms were not calculated.
However, as we have seen, as far as the $N_1$-dominated scenario is concerned,
phantom terms can be safely neglected in the calculation of
the final asymmetry, and therefore there is no contradiction
between our results and previous ones where phantom terms have not been 
identified \cite{riottodesimone,beneke}.

On the other hand, we are interested in accounting for heavy neutrino
flavour effects. On this case we cannot neglect the phantom terms since in this case,
as we discuss in the next Section, they can contribute to the total final asymmetry
and even dominate (phantom leptogenesis~\cite{phantom}).

%It is also quite interesting to notice that, even in the $N_1$-dominated scenario,
%the phantom terms can produce lepton flavour asymmetries larger than
%the observed baryon asymmetry.
%These cancelling flavour asymmetries can be as large as $10^{-5}-10^{-4}$
%for initial thermal $N_1$ abundance and in this case they could have potential effects
%at lower temperatures $T\lesssim 100\,{\rm MeV}$ when the
%thermal bath is flavour sensitive.  They can very unlikely have
%direct effects on primordial nuclear abundances since an impact on BBN
%requires much larger asymmetries  ($\gtrsim 0.01$ \cite{raffelt}).
%However, asymmetries as small as $\sim 10^{-5}$ could for example
%be relevant for active-sterile neutrino oscillations \cite{activesterile}.

%%%%%%%%%%%%%%%%%%%%%%%%%%%%%%%%%%%
\section{A simplified case with two charged lepton flavours}
%%%%%%%%%%%%%%%%%%%%%%%%%%%%%%%%%%%

In this section we account for heavy neutrino flavour effects
considering a simplified two charged lepton flavour case.
This will greatly simplify the notation making the new results more
easily readable.
It will be then quite straightforward in the next Section to generalise all the equations
to a realistic three lepton flavour case.

For definiteness we consider masses $M_i \gg 10^{9}\,{\rm GeV}$,
when only tauon lepton interactions have to be taken into account.
We also assume  that the heaviest RH neutrinos $N_3$ do not contribute to the final asymmetry.
This is in any case a valid assumption if $M_3\gg T_{RH}\gg M_2$, since in this way the  $N_3$'s
would not thermalise. As for lepton flavours,  we
will extend the results to the three heavy
neutrino flavour case in the next Section.

Notice that with these assumptions, the two charged lepton flavour case can
be regarded as a special  case where the two heavy neutrino lepton flavours, ${\ell}_1$ and ${\ell}_2$,
lie on the same plane  orthogonal to the $e-\mu$ plane and therefore $\tau_2^{\bot}=\tau_1^{\bot}=\tau^{\bot}$
(see Fig.~3).
\begin{figure}
\begin{center}
     \hspace*{10mm}
     \psfig{file=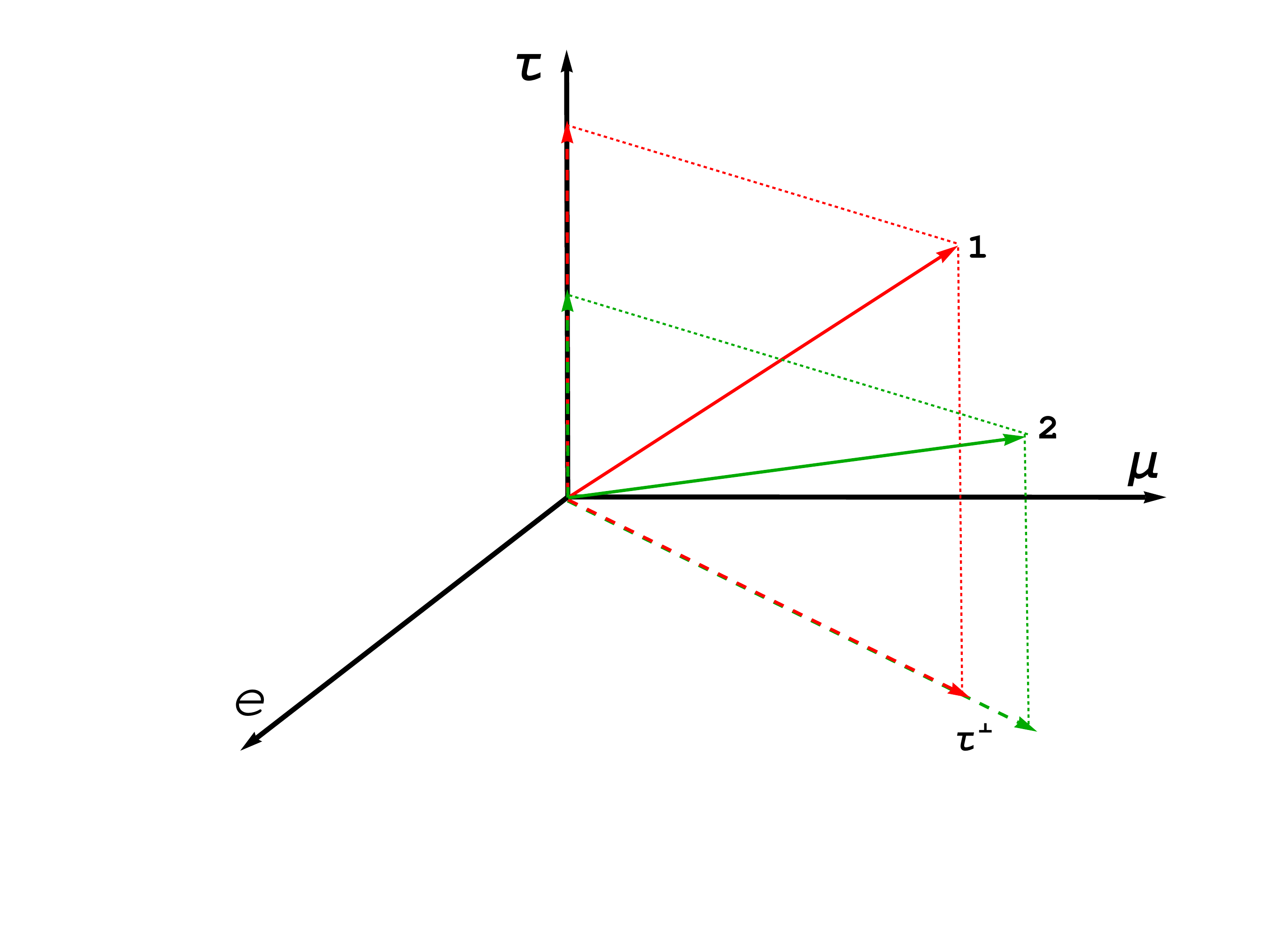,height=80mm,width=100mm}
     \vspace*{-10mm}
     \caption{Flavour configuration of the two heavy neutrino lepton flavours, ${\ell}_1$  and ${\ell}_2$,
                leading to the simplified two charged lepton neutrino flavour case considered in this section.}
\end{center}
\end{figure}
Correspondingly the two anti-lepton flavours,  $\bar{\ell}_1$ and $\bar{\ell}_2$,
also lie on the same plane  orthogonal to the $e-\mu$ plane and therefore
$\bar{\tau}_2^{\bot}=\bar{\tau}_1^{\bot} \equiv \bar{\tau}^{\bot}$ with
$\bar{\tau}^{\bot}$ that is now assumed to be $C\!P$ conjugated of $\tau^{\bot}$.
 In this way, in the whole following discussion in this Section, we
 will have only two charged lepton flavours, $\tau$ and $\tau^{\bot}$.

The density matrix equation Eq.~(\ref{fullyflavoured}), valid for $N_1$ leptogenesis,
gets then generalised into $(\a, \b= \tau, \tau^{\bot})$
\bea\label{denmaeqtwo}
{dN^{B-L}_{\a\b} \over dz}   & = &
\ve^{(1)}_{\a\b}\,D_1\,(N_{N_1}-N_{N_1}^{\rm eq})-{1\over 2}\,W_1\,\left\{{\cal P}^{(1)0}, N^{B-L}\right\}_{\a\b} \\ \nonumber
& + & \ve^{(2)}_{\a\b}\,D_2\,(N_{N_2}-N_{N_2}^{\rm eq})-{1\over 2}\,W_2\,\left\{{\cal P}^{(2)0}, N^{B-L}\right\}_{\a\b}
\\  \nonumber
& - &  {{\rm Im}(\L_{\t})\over H\, z} (\sigma_1)_{\a\b}\,N^{B-L}_{\a\b} \  ,
\eea
where $N_{N_2}$ is described, as $N_{N_1}$, by an analogous eq.~(\ref{dlg1}).
We now discuss the three asymptotic cases, the first one for
$M_2 \gg 10^{12}\,{\rm GeV} \gg M_1$, the
second for $M_2, M_1 \gg 10^{12}\,{\rm GeV}$
and the third for  $M_1, M_2 \ll 10^{12}\,{\rm GeV}$,
where Boltzmann equations are recovered.
In this way we will derive, within a density matrix formalism,
results that were already obtained within an instantaneous
collapse of the quantum state formalism:
the first is phantom leptogenesis \cite{phantom},
the second is the heavy neutrino flavour projection \cite{bcst,nir}.

\subsection{Case $M_2 \gg 10^{12}\,{\rm GeV} \gg M_1$: three stages phantom leptogenesis}
%%%%%%%%%%%%%%%%%%%%%%%%%%%%%%%%%%%%%%%%%%%%%%%%%%%%%%%

Let us consider the asymmetry production from the $N_2$'s  at $T\sim M_2$.
This is basically described by the same equations that we wrote in the previous
section for the $N_1$-dominated scenario where now simply all quantities
have to be relabelled in a way that $1\rightarrow 2$.
Since charged lepton interactions are negligible, we can
use the Eq.~(\ref{unflavoured2}) for the calculation of $N^{B-L}_{\a\b}$,
that now, in the charged lepton flavoured basis, simply becomes
\be\label{NBmLabN2}
{dN^{B-L}_{\a\b} \over dz} =
\ve^{(2)}_{\a\b}\,D_2\,(N_{N_2}-N_{N_2}^{\rm eq})
- {1\over 2} \, W_2 \, \left\{{\mathcal{P}}^{(2)0}, N^{B-L}\right\}_{\a\b} \,  ,
\ee
with an obvious re-definition of all quantities  that  now refer to $N_2$.
At $T\simeq T_{B2}\equiv M_2/z_{B2}$, the $\tau$ and $\tau^{\bot}$
asymmetries are described by the Eqs.~(\ref{NBmLttTB2}) with $1\rightarrow 2$,
\bea\label{NBmLttTB2}
N^{B-L}_{\t\t}(T\simeq T_{B2}) &\simeq& p^0_{2\t}\,N_{B-L}^{T\simeq T_{B2}} - {\D p_{2\t}\over 2} \, \k (K_2/2) \,  ,
\\ \nonumber
N^{B-L}_{\tau^{\bot}\tau^{\bot}}(T\simeq T_{B2}) &\simeq&  p^0_{2\tau^{\bot}}\,N_{B-L}^{T\simeq T_{B2}} +
{\D p_{2\t}\over 2}\, \k(K_2/2) \, .
\eea
Again, at temperatures below  $T_{\star}\ll  10^{12}\, {\rm GeV}$,
the $N^{B-L}_{\a\b}$ off-diagonal  terms are fully damped by the
tauon charged interactions,  so that the $N_{\Delta_{\tau}}$ and
$N_{\Delta_{\tau^{\bot}}}$ asymmetries, corresponding to the diagonal terms,
can be treated as measured quantities.

At $T\sim T_{B2}$, the phantom terms in the eq.~(\ref{NBmLttTB2})
cancel with each other and they do not contribute to the total asymmetry.
Therefore, so far, the description of the asymmetry evolution
is completely analogous to that one discussed in the $N_1$-dominated scenario.

However, there is still a third stage to be taken into account: the lightest RH neutrino washout.
For $T\sim M_1$, the tauon and the $\tau^{\bot}$ asymmetries
are washed out by the lightest RH neutrino inverse processes.
At $T\simeq T_{B1} = M_1/z_{B1}$, they get frozen to their final values
\bea\label{NBmLttphantom}
N_{\Delta \t}^{\rm f} &\simeq&
\left[p^0_{2\t}\,N_{B-L}^{T\simeq T_{B2}} - {\D p_{2\t}\over 2} \, \k (K_2/2) \right]\,e^{-{3\pi\over 8}\,K_{1\tau}} \,  ,
\\
N_{\D \tau^{\bot}}^{\rm f} &\simeq&  \left[ p^0_{2\tau^{\bot}}\,N_{B-L}^{T\simeq T_{B2}} +
{\D p_{2\t}\over 2}\, \k (K_2/2)  \right]\,e^{-{3\pi\over 8}\,K_{1\tau^{\bot}}} \,  ,
\eea
so that the final total asymmetry $N_{B-L}^{\rm f} \simeq N_{\Delta \t}^{\rm f} + N_{\D \tau^{\bot}}^{\rm f} $.
If  for the flavour $\a=\tau$ ($\tau^{\bot}$) one has    $K_{1\a}\lesssim 1$, while for the other
flavour $\beta=\tau^{\bot}$ ($\tau$)  one has $K_{1\beta} \gg 1$, the final asymmetry will be
dominated by the $\a$ asymmetry,
\be
N_{B-L}^{\rm f} \simeq
p^0_{2\a}\,N_{B-L}^{T\simeq T_{B2}} - {\D p_{2\a}\over 2} \, \k (K_2/2)  \, .
\ee
Interestingly the phantom term is affected by a washout at the production that is half that one
acting on the final asymmetry. Since in the strong wash-out regime approximately $\k(K_2)\propto 1/K_2^{1.2}$,
the phantom term contribution gets enhanced by a factor $3$ compared to that one
proportional to the total asymmetry at the production and this could make it dominant.  
Having included the effect of gauge interactions, now in the strong wash-out regime ($K_2 \gg 1$)
the phantom terms are independent of the initial conditions.
Therefore, phantom terms have to be included even in the case of initial vanishing abundance.

Phantom leptogenesis was first discussed within an instantaneous quantum state
collapse description without gauge interactions \cite{phantom}.
Here we have re-derived it within a density matrix formalism showing the importance of gauge interactions
that determine a wash-out of the phantom terms, though halved. Notice that
there are three well separated stages: $N_2$ asymmetry production at $T\simeq T_{B2}$, decoherence
at $T \sim T_{\star}$ and flavour asymmetric $N_1$ washout at $T\sim M_1$.

Notice also that phantom leptogenesis has some analogies with the
scenario of $N_1$-leptogenesis with $\ve_1=0$ \cite{nardi} that we discussed
in the previous section. In both cases the final asymmetry
originate from the $C\!P$ violating terms $\propto \Delta p_{i\a}$ due to a
different flavour composition of leptons and anti-leptons. In both cases
a non-vanishing final asymmetry relies on an asymmetric washout acting
on the two flavour asymmetries.
There are however important differences.
In the  case of $N_1$ leptogenesis with $\ve_1=0$ one has that production, decoherence
and washout occur simultaneously, while in the case of phantom leptogenesis they occur at different stages
and between the production and the $N_1$ washout  stage the phantom terms they
cancel in the final asymmetry.
Another important difference is that in the case of phantom leptogenesis one has not
to assume the special assumption $\ve_1$ or $\ve_2=0$ ($B-L$ conservation):
 if the washout at the production is sufficiently strong
 phantom terms can potentially dominate because of the reduced wash-out compared to
 the total asymmetry.

 As we are going to show, phantom leptogenesis is even
 more general and it does not necessarily require that the $N_2$ production and
 the $N_1$  washout stages occur in two different fully flavoured regimes.

\subsection{Case $M_2 \gtrsim 3\, M_1 \gg 10^{12}\,{\rm GeV}$:  heavy neutrino flavour projection
and two stages phantom leptogenesis}
%%%%%%%%%%%%%%%%%%%%%%%%%%%%%%%%%%%%%%%%%%%%%%%%%%%%%

Let us now consider the case when both heavy neutrino masses
$M_2, M_1 \gg  10^{12}\,{\rm GeV}$ and charged lepton interactions
do not affect the final asymmetry.
This can be called the heavy flavoured scenario \cite{problem} since the only lepton flavours that
affect the final asymmetry are those produced from the heavy RH neutrinos.
The density matrix equation (\ref{denmaeqtwo}) can then be recast simply  as
\bea\label{denmaeqheavyfl}
{dN^{B-L}_{\a\b} \over dz}   & = &
\ve^{(1)}_{\a\b}\,D_1\,(N_{N_1}-N_{N_1}^{\rm eq})-{1\over 2}\,W_1\,\left\{{\cal P}^{(1)0}, N^{B-L}\right\}_{\a\b} \\ \nonumber
& + & \ve^{(2)}_{\a\b}\,D_2\,(N_{N_2}-N_{N_2}^{\rm eq})-{1\over 2}\,W_2\,\left\{{\cal P}^{(2)0}, N^{B-L}\right\}_{\a\b} \, . \eea

\subsubsection{Projection effect (in isolation)}

For illustrative purposes, we want first to describe just the $N_1$ washout
of the asymmetry produced by the $N_2$ decays, without any additional effect.
Therefore, we first neglect  the different flavour compositions of
leptons and anti-leptons assuming that $\D p_{1\a} = \D p_{2\a} = 0$.

With such a simplifying assumption, the lepton quantum states are given by
\be
| 1\rangle  =
{\cal C}_{1\tau}\,|\tau \rangle + {\cal C}_{1\tau^{\bot}}\,|\tau^{\bot} \rangle \,
\hspace{5mm}\mbox{\rm and}\hspace{5mm}
| \bar{1}\rangle  =
{\cal C}^{\star}_{1\tau}\,|\bar{\tau} \rangle +
{\cal C}^{\star}_{1\tau^{\bot}}\,|\bar{\tau}^{\bot} \rangle  \;\;\; (C\! P| \bar{1}\rangle =| 1\rangle)\, ,
\ee
\be
| 2\rangle  =
{\cal C}_{2\tau}\,|\tau \rangle + {\cal C}_{2\tau^{\bot}}\,|\tau^{\bot} \rangle \,
\hspace{5mm}\mbox{\rm and}\hspace{5mm}
|\bar{2}\rangle  =   {\cal C}^{\star}_{2\tau}\,|\bar{\tau} \rangle +
{\cal C}^{\star}_{2 \tau^{\bot}}\,|\bar{\tau}^{\bot} \rangle \;\;\; (C\! P| \bar{2}\rangle =| 2\rangle) \, .
\ee
Assuming  the hierarchical limit, $M_2\gtrsim 3\,M_1$ \cite{hierarchical},
there are two well distinguished different stages.
In a  first stage at $T\sim M_2$, an asymmetry is produced from $N_2$ decays.
The lepton density matrix is then given by $\rho^{\ell}_{ij}={\rm diag}(1,0)$
in the basis ${\ell_2-\ell_2^{\bot}}$.  Analogously the anti-lepton density matrix
is  given by $\rho^{\bar{\ell}}_{i j}={\rm diag}(1,0)$
 in the basis $\bar{\ell}_2-\bar{\ell_2}^{\bot}$ that at the moment we are assuming to
 be  $C\!P$ conjugated of  ${\ell_2-\ell_2^{\bot}}$. As in the previous subsection,
 the asymmetry production from $N_2$ decays is again described by the Eq.~(\ref{NBmLabN2})
with vanishing phantom terms so that we simply have
 \be\label{NBmLttTB2bis}
N^{B-L}_{\t\t}(T\simeq T_{B2}) \simeq  p^0_{2\t}\,N_{B-L}^{T\simeq T_{B2}} \,  ,
\hspace{5mm}
N^{B-L}_{\tau^{\bot}\tau^{\bot}}(T\simeq T_{B2})  \simeq  p^0_{2\tau^{\bot}}\,N_{B-L}^{T\simeq T_{B2}} \, ,
\ee
where $N_{B-L} ^{T \simeq T_{B2}} \simeq \ve_2\,\k(K_2)$.
We have now to consider the $N_1$ washout stage at $T\sim M_1$.  Since at the moment
we are just interested in describing the $N_1$ washout, we also neglect the
$N_1$ asymmetry production assuming a vanishing $\ve_{\a\b}^{(1)}$.
Moreover let us first further assume, just for simplicity,
$|1\rangle = |\tau \rangle$ and correspondingly $|\bar{1}\rangle = |\bar{\tau}\rangle$.

In this way, at $T\sim M_1$, the Eqs.~(\ref{denmaeqheavyfl}) for the asymmetry
evolution in the charged lepton flavour basis can be simply rearranged as ($\a, \b = \tau, \tau^{\bot}$)
 \be\label{denmaeqtwoN1washout}
{dN^{B-L}_{\a\b} \over dz} =  - W_1\,\left(\begin{array}{cc}
N^{B-L}_{\t\t} & {1\over 2}\,N^{B-L}_{\t\t^{\bot}} \\
{1\over 2}\,N^{B-L}_{\t^{\bot}\t} & 0
\end{array}\right)  \, ,
\ee
and, at the end of the $N_1$-washout at $T\simeq T_{B1}$, one simply finds
\be\label{NBmLttTB1bis}
N^{B-L}_{\t\t}(T\simeq T_{B1}) \simeq
e^{-{3\pi\over 8}\,K_1}\,p^0_{2\t}\,N_{B-L}^{T\simeq T_{B2}} \,  ,
\hspace{5mm}
N^{B-L}_{\tau^{\bot}\tau^{\bot}}(T\simeq T_{B1})  \simeq  p^0_{2\tau^{\bot}}\,N_{B-L}^{T\simeq T_{B2}} \,  .
\ee
Finally, at $T\sim 10^{12}\,{\rm GeV}$, the
charged lepton interactions  damp the off-diagonal
terms measuring the tauon and the `non-tauon' (i.e. the $\tau^{\bot}$)
asymmetries.

This result can be easily generalised. Let us, first of all, allow
an arbitrary $|1\rangle$ flavour composition but continuing, for the time being,
to neglect the $N_1$ asymmetry production, at $T\sim T_{B1}$. The Eq.~(\ref{denmaeqtwoN1washout})
has now to be written in the basis ${\ell}_1-{\ell}_1^{\bot}$,
\be\label{denmaeqtwoN1washoutgeneral}
{dN^{B-L}_{i_1 j_1} \over dz} =  - W_1\,
\left(\begin{array}{cc}
N^{B-L}_{1 1} & {1\over 2}\,N^{B-L}_{1 1^{\bot}} \\
{1\over 2}\,N^{B-L}_{1^{\bot} 1} & 0
\end{array}\right) \hspace{10mm} (i_1, j_1 = 1, 1^{\bot})
\, .
\ee
The solution is  again quite trivial in this basis: the $11$ term
is washed out,
\be
N^{B-L}_{11}(T \simeq T_{B1}) = e^{-{3\pi\over 8}\, K_1}\, N^{B-L}_{11}(T \simeq T_{B2}) \, ,
\ee
together with the off-diagonal terms,
while the $1^{\bot}1^{\bot}$ term  is unwashed.
The asymmetry matrix at $T\sim T_{B2}$,
in the ${\ell}_1-{\ell}_1^{\bot}$ basis,  can now be calculated in
terms of the rotation matrices (cf. Eq.~(\ref{rotationmatrices})) as
\be
N^{B-L}_{i_1 j_1}(T \simeq T_{B2}) = N_{B-L}^{T\simeq T_{B2}}\,
R^{(1)0\dagger}_{i_1 \a } \, R^{(2)0}_{\a i_2}
\,
\left(\begin{array}{cc}
1 & 0 \\
0 & 0
\end{array}\right)
\,
\, R^{(2)0\dagger}_{j_2 \b} \, R^{(1)0}_{\b j_1} \, .
\ee
In a more compact way, considering that
$N^{B-L}(T\simeq T_{B2}) = N_{B-L}^{T\simeq T_{B2}} |2\rangle\langle 2|$,
this can be more conveniently written as
\be
N^{B-L}_{i_1 j_1}(T \simeq T_{B2}) =
N_{B-L}^{T\simeq T_{B2}}\, \left(\begin{array}{cc}
p_{12} & \langle 1 | 2 \rangle \langle 2 | 1^{\bot}\rangle \\
\langle 1^{\bot} | 2 \rangle \langle 2 | 1 \rangle & 1-p_{12}
\end{array}\right)  \, ,
\ee
where \cite{problem}
\be
p_{12} \equiv | \langle {\ell_1} | {\ell_2} \rangle |^2 =
{\left|(h^{\dagger}\, h)_{12}\right|^2
\over (h^{\dagger} \, h)_{11}\,(h^{\dagger} \, h)_{22}} \, .
\ee
The final asymmetry can then be calculated as
\be\label{NBmLfsec3}
N_{B-L}^{\rm f} =  {\rm Tr}[N^{B-L}_{i_1 j_1}(T \simeq T_{B1})] =
e^{-{3\pi\over 8}\, K_1}\, p_{12} \, N_{B-L} ^{T \simeq T_{B2}} +
(1-p_{12}) \, N_{B-L}^{T \simeq T_{B2}} \, .
\ee
The asymmetry can be also rotated in the charged lepton flavour basis,
\be
N^{B-L}_{\a\b} (T \simeq T_{B1}) = R^{(1)0}_{\a i_1} \, N^{B-L}_{i_1 j_1}(T \simeq T_{B1}) \, R^{(1)0\dagger}_{j_1 \b} \, .
\ee
At $T \simeq 10^{12}\,{\rm GeV}$ the charged lepton interactions
just damp the off-diagonal terms without affecting the  total asymmetry
given by the trace and for this reason we could directly write  the Eq.~(\ref{NBmLfsec3}).

This result  fully confirms  what one expects
 within an instantaneous quantum state collapse
description.
It is not only confirmed that just the ${\ell}_1$-parallel component of the asymmetry
undergoes the $N_1$ washout  while the  orthogonal component  completely escapes it \cite{bcst,nir},
but also that the washout of the parallel component is exactly described by the factor
$\exp[-(3\pi K_1 / 8)]$, independently of the value of $K_1$  \cite{problem}.
Notice that in an intermediate regime $K_1\sim 1$,
the quantum states at $T \simeq T_{B1}$ are left in a sort of partially incoherent
mixture, with some residual flavour oscillations
that however do not affect the total asymmetry.

Notice that this result also applies to a
possible pre-existing asymmetry produced by some
other external mechanism \cite{nir,problem}.
Therefore, the conclusions of \cite{problem},  employing
this result in various situations, are also confirmed.

One can then easily further generalise this result  accounting also for
a possible $N_1$ asymmetry generation, simply obtaining for the final asymmetry
\be\label{NBmLfsec3b}
N_{B-L}^{\rm f} =
\ve_1\,\k(K_1) + \left(e^{-{3\pi\over 8}\, K_1}\, p_{12} + 1-p_{12} \right) \, \ve_2\,\k(K_2) \, .
\ee

\subsubsection{Projection effect in combination with phantom leptogenesis}

We still miss a last step. We have so far
assumed that the flavour compositions of the ${\ell}_2$
and ($C\!P$ conjugated) $\bar{\ell}_2$ quantum states  are  the same.
We want now to show that, when this additional
flavoured $C\!P$ violation contribution is taken into account,
phantom terms contribute to the final asymmetry and the
eq.~(\ref{NBmLfsec3b}) gets generalised.
Notice that this time the role played by the charged lepton flavour basis in the previous
subsection, is replaced by the heavy neutrino lepton basis ${\ell}_1$--${\ell}_1^{\bot}$.
Notice that in general now also the  basis ${\ell}_1-{\ell}_1^{\bot}$
does not coincide with $\bar{\ell}_1-\bar{\ell}_1^{\bot}$ and therefore there
can be an ambiguity about the basis on which one should project. However,
one can  calculate the wash-out in the tree-level basis $1^0-1^{0\bot}$,
so that the eq.~(\ref{NBmLabN1}) can be still used also in this case.

Therefore, the  quantum states $|2\rangle$ and  $|\bar{2}\rangle$
have now to be projected, more generally, on the tree-level basis $1^0-1^{0\bot}$
so that they can be written as
\be
| 2\rangle  =
\langle 1^0 | 2 \rangle \,|1^0 \rangle + \langle  1^{0\bot} | 2 \rangle  \,|1^{0\bot} \rangle \,
\hspace{5mm}\mbox{\rm and}
\hspace{5mm}
| \bar{2} \rangle  =   \langle \bar{1}^0 | \bar{2} \rangle \,|\bar{1}^0 \rangle +
\langle  \bar{1}^{0\bot} | \bar{2} \rangle \,|\bar{1}^{0\bot} \rangle \, .
\ee

Therefore,  writing the Eq.~(\ref{NBmLabN2}) in this basis, we have at the production
($i_1^{0},j_1^{0} = 1^0, 1^{0\bot}$)
\be\label{NBmLabN2i1j1}
{dN^{B-L}_{i_1^0 j_1^0} \over dz} =
\ve^{(2)}_{i_1^0 j_1^0}\,D_2\,(N_{N_2}-N_{N_2}^{\rm eq})
- {1\over 2} \, W_2 \, \left\{{\mathcal{P}}^{(2)0}, N^{B-L}\right\}_{i_1^0 j_1^0} \,  ,
\ee
where as usual the superscript ``$0$'' indicates the tree level quantities
that can be approximately fully employed in the calculation of the washout term. In this way
we obtain expressions for the heavy neutrino lepton flavour asymmetries,
that are analogous  to the eqs.~(\ref{NBmLttTB2}-\ref{NBmLttTB2bis}) for the charged lepton flavoured asymmetries,
\bea\label{NBmL11TB2}
N^{B-L}_{1^0 1^0}(T\simeq T_{B2}) &\simeq& p^0_{12}\,\ve_2\,\k(K_2) - {\D p_{21^0}\over 2} \, \k(K_2/2) \,  ,
\\
N^{B-L}_{1^{0\bot}1^{0\bot}}(T\simeq T_{B2}) &\simeq&  (1-p^0_{12})\,\ve_2\,\k(K_2)
+  {\D p_{21^0}\over 2}\, \k(K_2/2) \, .
\eea
The quantity $\Delta p_{21^0}$ is defined analogously to the $\D p_{i\a}$'s (cf. eqs.~(\ref{deltap1alpha}), (\ref{epsial})),
explicitly $\Delta p_{21^0}\equiv |\langle 1^0 | 2 \rangle|^2 - |\langle \bar{1}^0 | \bar{2} \rangle|^2$.
Finally, taking into account the lightest RH neutrino washout and asymmetry production,
we obtain for the final asymmetry
\be
N_{B-L}^{\rm f} =
\ve_1\,\k(K_1) + \left[p^0_{12} \, e^{-{3\pi\over 8}\, K_1}
+ (1-p^0_{12})  \right] \, \ve_2\,\k(K_2)  +
\left(1-e^{-{3\pi\over 8}\, K_1}\right) \, {\Delta p_{21^0}\over 2}\, \k(K_2/2)\, .
\ee
Therefore, the phantom terms give an  additional contribution to both components
and in particular to the orthogonal component.
If $K_1 \ll 1$, both the parallel and the orthogonal
components are unwashed and the phantom terms cancel with each other.
On the other hand, in the opposite case, for $K_1 \gg 1$, the parallel component is
completely washed out so that only the orthogonal one survives (together with the
additional $N_1$-unwashed phantom term contribution).

This result shows that phantom leptogenesis goes even beyond
the case where the two RH neutrino
masses fall into two different flavour regimes \cite{phantom}.

Finally, it should be clear that an account of the
different flavour compositions of
 the ${\ell}_1$   and $\bar{\ell}_1$ quantum states at the production  from $N_1$,
would lead to additional phantom terms.  These, however, cancel with
each other and do not contribute to the final asymmetry,
as already discussed in section 2.

\subsection{Case $10^{12}\,{\rm GeV} \gg M_1, M_2$}
%%%%%%%%%%%%%%%%%%%%%%%%%%%%%%%%

When $M_1, M_2 \ll 10^{12}\,{\rm GeV}$ both RH neutrinos produce their asymmetry
 in the two-flavour regime.
The production from the heavier RH neutrinos is given by the usual result
%valid in the
%strong washout regime for both flavor $K_{2{\tau}^{\bot}}, K_{2\tau} \gg 1$,
\be\label{fullystrong}
N^{B-L}_{\t\t} (T\simeq T_{B2}) = \ve_{2\t} \, \k(K_{2{\tau}}) \, ,\hspace{1cm} N^{B-L}_{\t^{\bot}\t^{\bot}} (T\simeq T_{B2})
= \ve_{2\t^{\bot}} \, \k(K_{2{\tau}^{\bot}}) \, .
\ee
In the strong washout regime for both flavours, $K_{2{\tau}^{\bot}}, K_{2\tau} \gg 1$,
the sum, i.e. the total asymmetry, can be approximated by the  Eq.~(\ref{twofully}) rewritten for the heavier RH neutrino.
When the temperature drops down to $T\sim T_{B1}$, the washout from the lighter RH neutrino starts to
act. Similarly to the previous cases, this washout factorizes from the general expression and can be
expressed as a simple exponential pre-factor so that
\bea\label{ciccio}
N^{B-L}_{\t\t} (T\simeq T_{B2}) &=&  \ve_{2\t} \, \k(K_{2{\tau}})\, e^{-{3\pi\over 8}\,K_{1\t} } \, , \\
N^{B-L}_{\t^{\bot}\t^{\bot}} (T\simeq T_{B2}) &=&
\ve_{2\t^{\bot}} \, \k(K_{2{\tau}^{\bot}}) \,
e^{-{3\pi\over 8}\,K_{1\t^{\bot}} }\, .
\eea
The production of the asymmetry from the $N_1$ decays is then added to
what is left from the $N_2$ production, so that we finally obtain
\bea\label{}
N^{B-L}_{\t\t} (T\simeq T_{B1}) &=&\ve_{2\t} \, \k(K_{2{\tau}})\, e^{-{3\pi\over 8}\,K_{1\t} }
+ \ve_{1\t} \, \k(K_{1{\tau}})\, , \\
N^{B-L}_{\t^{\bot}\t^{\bot}} (T\simeq T_{B1}) &=&
\ve_{2\t^{\bot}} \, \k(K_{2{\tau}^{\bot}}) \,
e^{-{3\pi\over 8}\,K_{1\t^{\bot}} } + \ve_{1\t^{\bot}} \, \k(K_{1{\tau}^{\bot}}) \, .
\eea
It should be noticed that there are no phantom terms in this case because of the assumption made at the beginning of the Section that $\tau_2^{\bot}= \tau_1^{\bot} \equiv \tau^{\bot}$. In this
case, we have an effective two-flavour problem and there are no 
phantom terms cancelling out. If we relax
the two-flavour  assumption allowing $\tau_2^{\bot}\neq \tau_1^{\bot} $, we have to work
 in a full three-flavour basis, and, as we will see, phantom terms appear again
 in the final asymmetry.
 %  it is expected that part
%of the asymmetry, the phantom terms,
%will escape the washout at the $N_2$ production, and an additional asymmetry
%(still dependent on the initial $N_2$ abundance) will emerge from an
%asymmetric washout by $N_1$.
We discuss this more general case in the next Section.

%%%%%%%%%%%%%%%%%%%%%%%%%%%%%%%%%%%%%%%%%%%%%%%%%%%%%%%
\section{General case with three charged lepton flavours and three heavy neutrino flavours}
%%%%%%%%%%%%%%%%%%%%%%%%%%%%%%%%%%%%%%%%%%%%%%%%%%%%%%%

If we consider the general realistic case with
three lepton flavours,
the density matrix equations have to be written in terms of $3\times 3$ matrices.  In
general the three heavy neutrino flavours have no particular flavour orientations
in the three charged lepton flavour space (see Fig.~4).
\begin{figure}
\begin{center}
     \hspace*{10mm}
     \psfig{file=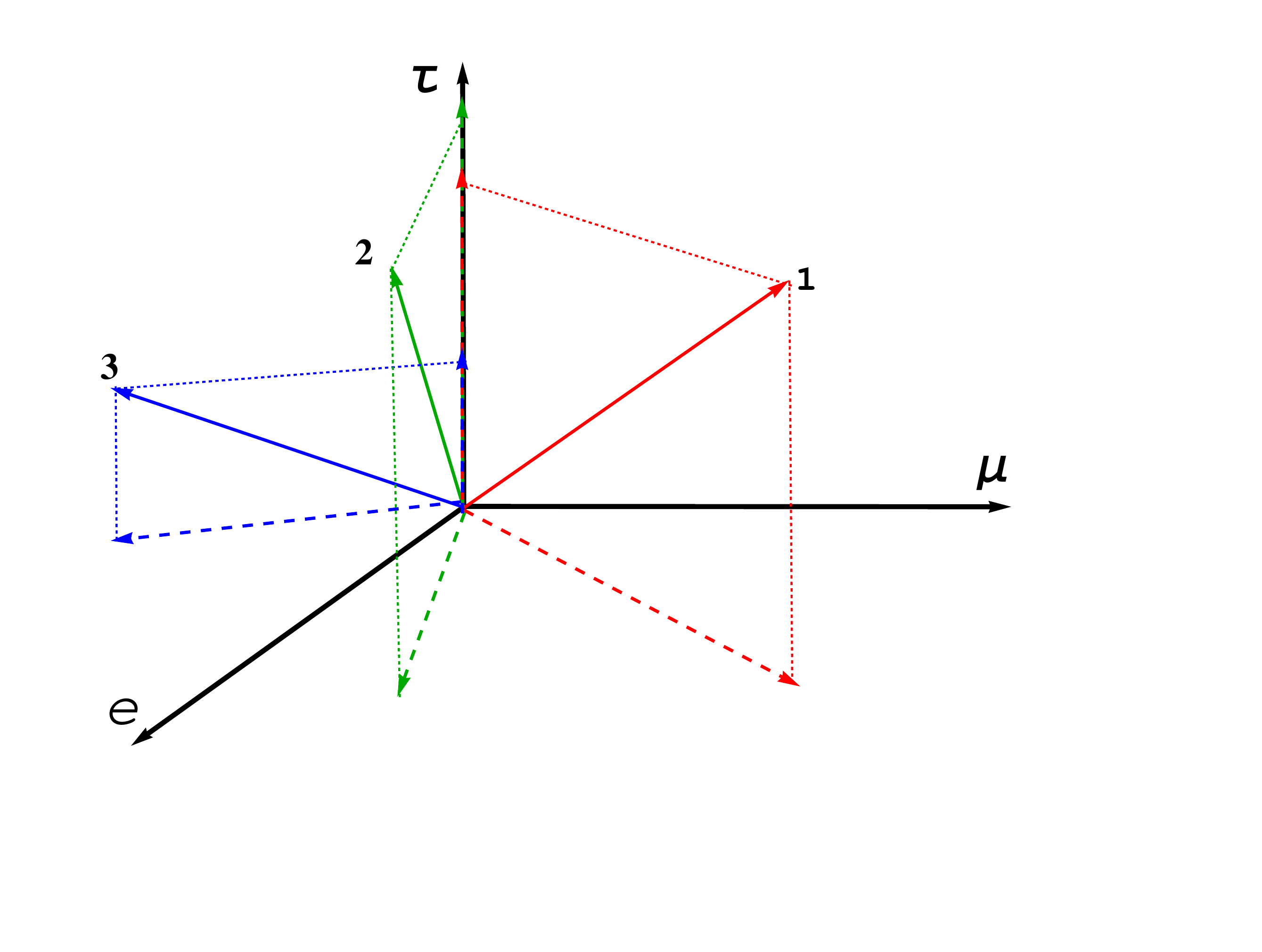,height=90mm,width=105mm}
     \vspace*{-15mm}
     \caption{A generic three heavy neutrino lepton flavour configuration.}
\end{center}
\end{figure}
If we also consider generic three RH neutrinos mass patterns with masses $M_i \gg 10^6 \,{\rm GeV}$,
the density matrix equation eq.~(\ref{denmaeqtwo})
further generalises into ($\a,\b=\t,\m, e$)
\bea\label{denmaeqfinal}
{dN^{B-L}_{\a\b} \over dz}   & = & \ve^{(1)}_{\a\b}\,D_1\,(N_{N_1}-N_{N_1}^{\rm eq})-{1\over 2}\,W_1\,\left\{{\cal P}^{(1)0}, N^{B-L}\right\}_{\a\b} \\ \nonumber
& + & \ve^{(2)}_{\a\b}\,D_2\,(N_{N_2}-N_{N_2}^{\rm eq})-{1\over 2}\,W_2\,\left\{{\cal P}^{(2)0}, N^{B-L}\right\}_{\a\b}  \\
\nonumber
& + & \ve^{(3)}_{\a\b}\,D_3\,(N_{N_3}-N_{N_3}^{\rm eq})-{1\over 2}\,W_3\,\left\{{\cal P}^{(3)0}, N^{B-L}\right\}_{\a\b} \\
& - &  {\rm Im}(\L_{\t})\left[\left(\begin{array}{ccc}
1 & 0 & 0 \\
0 & 0 & 0 \\
0 & 0 & 0
\end{array}\right),\left[\left(\begin{array}{ccc}
1 & 0 & 0 \\
0 & 0 & 0 \\
0 & 0 & 0
\end{array}\right),N^{B-L} \right]\right]_{\a\b}  \\ \nonumber
& - &
{\rm Im}(\L_{\m})\left[\left(\begin{array}{ccc}
0 & 0 & 0 \\
0 & 1 & 0 \\
0 & 0 & 0
\end{array}\right),\left[\left(\begin{array}{ccc}
0 & 0 & 0 \\
0 & 1 & 0 \\
0 & 0 & 0
\end{array}\right),N^{B-L} \right]\right]_{\a\b}
 \,   .
\eea
We have implied the effect of gauge interactions
in setting the condition of thermal equilibrium on the lepton abundances.

If one of the three masses is lower than $\sim 10^{6}\,{\rm GeV}$, electron flavour interactions
terms have to be included as well, though they have no real
impact, within this framework,  on the final asymmetry. This is because the
electron asymmetry is in any case already measured as a
`neither-muon-nor-tauon' asymmetry.

This master equation can now be used to calculate the final asymmetry
not only for all the ten mass patterns shown in Fig.~1, but also when
the $M_i$'s fall in one of the flavour transition regimes.

Notice that, though in this paper we are only considering hierarchical
RH neutrino mass patterns, this equation can also be used to calculate the asymmetry
beyond the hierarchical limit \cite{hierarchical} and even in the resonant case \cite{resonant}.
In this latter case, however,  many different  effects
can become important and should be included \cite{subtleresonant}.

Solutions of this set of equations are particularly difficult when at least two of the five
kinds of interactions are simultaneously effective, something that goes beyond
our objectives.
Here, as an example with three flavours, we want to show a particularly
interesting asymptotic limit  that  cannot be described  within the simplified
two-flavour case discussed in the previous section: the  two RH neutrino model  \cite{2RHneutrino}.
We will show that,  even in this case, phantom terms have in general
to be taken into account.

\subsection{Boltzmann equations for the two RH neutrino model}

We consider a  two RH neutrino model \cite{fgy} corresponding to a
situation where $M_3$ is sufficiently large ($M_3 \gg 10^{14}\,{\rm GeV}$)
to decouple in the seesaw formula for the calculation of the neutrino masses \cite{various}.
In order to reproduce the observed baryon asymmetry one has to impose
$M_1 \gtrsim 10^{9}\,{\rm GeV}$ so that the muon
interactions can be neglected in the Eq.~(\ref{denmaeqfinal})
On the other hand, in order to have $M_1$ and $M_2$ as low as possible,
it is interesting to consider the case
$10^{12}\,{\rm GeV} \gg M_2 \gtrsim 3\,M_1 \gg 3\times 10^{9}\,{\rm GeV}$
in a way to obtain a RH neutrino mass spectrum corresponding to the third panel
(from upper left) in Fig.~1.

This model has been recently revisited in \cite{2RHneutrino}.
We want here to re-derive, starting from the density matrix equation (\ref{denmaeqfinal}),
the Boltzmann kinetic equations  and the consequent formula for the final
asymmetry that in \cite{2RHneutrino} has been used to calculate the value of $M_1$
necessary to reproduce the observed baryon asymmetry
\footnote{It has been shown in \cite{2RHneutrino} that even the 
$N_2$ contribution to the asymmetry depends just on $M_1$ and not on $M_2$, provided that this
 is much smaller than $10^{12}\,{\rm GeV}.$}.

Thanks to the hierarchical limit,
we can again introduce different simplifications. First of all we can impose the
complete damping of the $\tau\alpha$ and $\alpha\tau$
($\alpha\neq\tau$) off-diagonal terms in the asymmetry matrix.

Second, we can consider the $N_2$ production at $T\simeq T_{B2}$.
With these assumptions, only the
$N_2$-terms can be considered in the Eq.~(\ref{denmaeqfinal})
and the asymmetry matrix can be  treated as a $2\times 2$ matrix in $\tau$--$\tau_2^{\bot}$ flavour space.
In this way the density matrix equation
reduce to a set of two Boltzmann equations
in an effective two fully flavoured regime,
\bea\label{dNBmLtautau}
{dN^{B-L}_{\t\t}\over dz} & = &
\ve^{(2)}_{\t\t}\,D_2\,(N_{N_2}-N_{N_2}^{\rm eq})-
p_{2\t}^{0}\,W_2\,N^{B-L}_{\t\t}  \, , \\ \label{dNBmLtaubot}
{dN^{B-L}_{\tau_2^{\bot}\tau_2^{\bot}}\over dz} & = &
\ve^{(2)}_{\tau_2^{\bot}\tau_2^{\bot}}\,D_2\,(N_{N_2}-N_{N_2}^{\rm eq})-
p_{2\tau_2^{\bot}}^{0}\,W_2\,N^{B-L}_{\t_2^{\bot}\t_2^{\bot}}   \, .
\eea
As usual, assuming first for simplicity that  the $|\tau_2^{\bot} \rangle$
 and the $|\bar{\tau}_2^{\bot} \rangle$ quantum states
 have the same flavour compositions, one finds
\be\label{ciccio2}
N^{B-L}_{\t\t} (T\simeq T_{B2}) = \ve_{2\t} \, \k(K_{2{\tau}})
\hspace{4mm}\mbox{\rm and} \hspace{4mm}
N^{B-L}_{\t_2^{\bot}\t_2^{\bot}} (T\simeq T_{B2})  =
\ve_{2\t_2^{\bot}} \, \k(K_{2{\tau}^{\bot}}) \, .
\ee
These values of the asymmetries at the end of the $N_2$ production stage
have to be used as initial values in the set of equations describing the
evolution of the asymmetries during the $N_1$ production,
 \bea\label{dNBmLtautau}
{dN^{B-L}_{\t\t}\over dz} & = &
\ve^{(1)}_{\t\t}\,D_1\,(N_{N_1}-N_{N_1}^{\rm eq})-
p_{1\t}^{0}\,W_1\,N^{B-L}_{\t\t}  \, , \\ \label{dNBmLtaubot}
{dN^{B-L}_{\tau_1^{\bot}\tau_1^{\bot}}\over dz} & = &
\ve^{(1)}_{\tau_1^{\bot}\tau_1^{\bot}}\,D_1\,(N_{N_1}-N_{N_1}^{\rm eq})-
p_{1\tau_1^{\bot}}^{0}\,W_1\,N^{B-L}_{\t_1^{\bot}\t_1^{\bot}}   \, .
\eea
The $\tau_2^{\bot}$ component of the asymmetry at the end of the
$N_2$ production has to be decomposed into a
$\t^{\bot}_1$ parallel component and into a $\t^{\bot}_1$
orthogonal component that we indicate with the symbol $\t^{\bot}_{1^{\bot}}$.
In this way one finds that the final asymmetry is the
sum of three flavour components (see Fig.~5),
\begin{figure}
\begin{center}
     \hspace*{10mm}
     \psfig{file=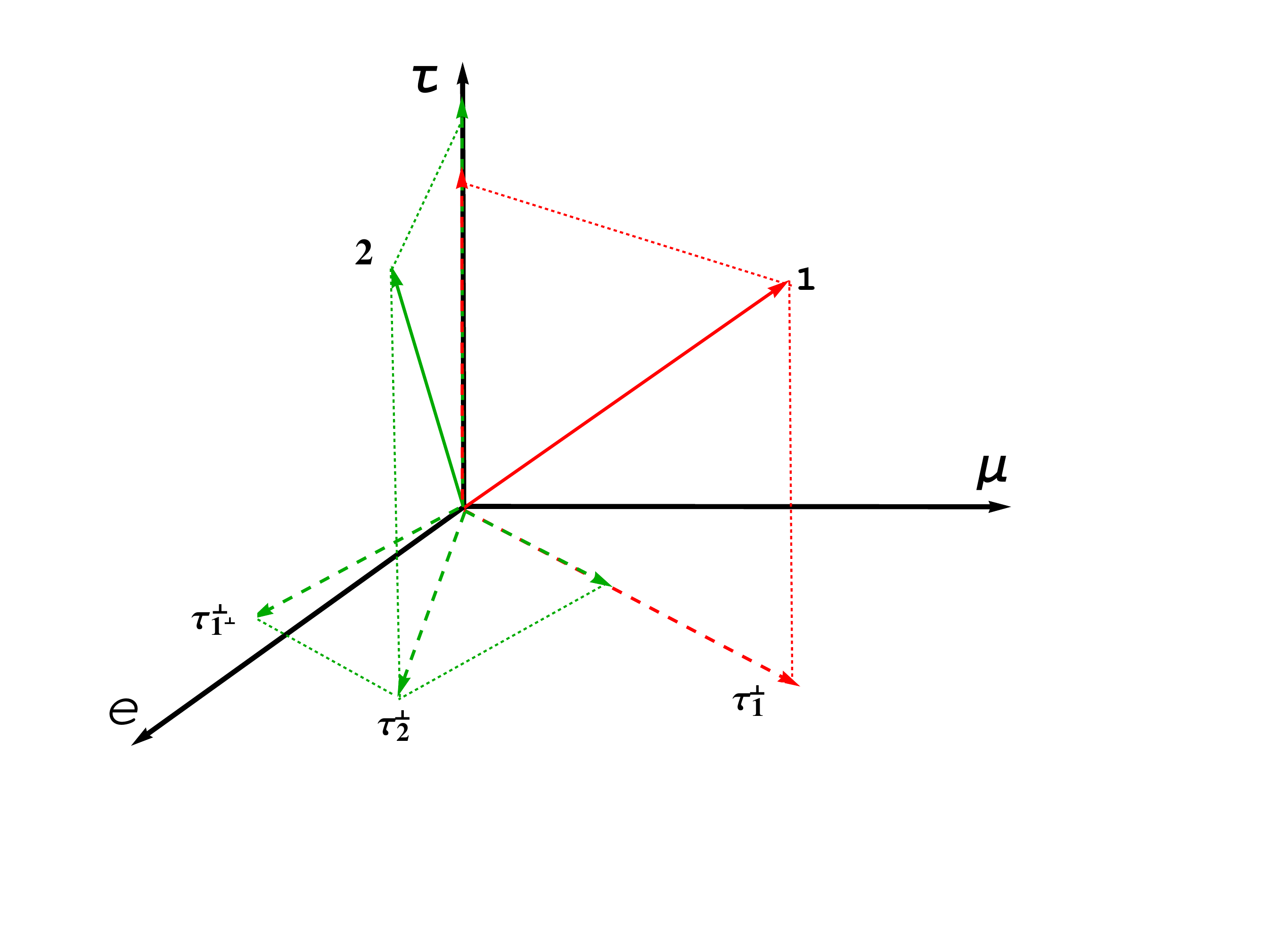,height=90mm,width=105mm}
     \vspace*{-10mm}
     \caption{Relevant lepton flavours in the two RH neutrino model.}
\end{center}
\end{figure}
\be
N^{\rm f}_{B-L}=
N^{B-L}_{\tau\tau}(T\simeq T_{B1}) +
N^{B-L}_{\tau_1^{\bot}\tau_1^{\bot}}(T\simeq T_{B1}) +
N^{B-L}_{\tau_{1^{\bot}}^{\bot}\tau_{1^{\bot}}^{\bot}}(T\simeq T_{B1}) \,  ,
\ee
where
\bea\label{ciccio4}
N^{B-L}_{\tau\tau}(T\simeq T_{B1}) & = & \ve_{1\t}\,\k(K_{1\t}) +  \ve_{2\t}\,\k(K_{2\t})\,e^{-{3\pi\over 8}\,K_{1\t}}  \,  ,\\
N^{B-L}_{\tau_1^{\bot}\tau_1^{\bot}}(T\simeq T_{B1})  & = &
\ve_{1\t_1^{\bot}}\,\k(K_{1\t_1^{\bot}}) +
  p^0_{\t_1^{\bot}\t_2^{\bot}}\ve_{2\t_2^{\bot}}\,\k(K_{2\t_2^{\bot}}) \,e^{-{3\pi\over 8}\,K_{1\t_1^{\bot}}}
\, ,\\
N^{B-L}_{\tau_{1^{\bot}}^{\bot}\tau_{1^{\bot}}^{\bot}}(T\simeq T_{B1}) & = &
\left(1-p^0_{\t_1^{\bot}\t_2^{\bot}}\right)\, \ve_{2\t_2^{\bot}}\,\k(K_{2\t_2^{\bot}})  \, .
\eea
This expression coincides with the result found in \cite{2RHneutrino} and is valid
neglecting phantom terms.
If one takes into account the different flavour compositions between the $|\tau_2^{\bot} \rangle$
 and the $|\bar{\tau}_2^{\bot} \rangle$ quantum states, then phantom terms are, in general, present.
 The procedure is essentially
 the same discussed in Section 3.2, with the only difference that now the phantom terms
 will appear only in the $\tau_1^{\bot}$ and $\tau_{1^{\bot}}^{\bot}$
 components but not in the measured tauon component. We can therefore directly
 write the final  result,
 \bea
 N_{B-L}^{\rm f} & = &   \ve_{1\t}\,\k(K_{1\t}) +  \ve_{2\t}\,\k(K_{2\t})\,e^{-{3\pi\over 8}\,K_{1\t}} \\  \nonumber
  & + &   \ve_{1\t_1^{\bot}}\,\k(K_{1\t_1^{\bot}}) +
  \left(p^0_{\t_1^{\bot}\t_2^{\bot}}\ve_{2\t_2^{\bot}}\,\k(K_{2\t_2^{\bot}}) -
  {\D p_{\t_2^{\bot}\t_{1^0}^{\bot}}\over 2}\,\k(K_{2\t_2^{\bot}}/2)  \right)
  \,e^{-{3\pi\over 8}\,K_{1\t_1^{\bot}}}         \\  \nonumber
 & + &  \left(1-p^0_{\t_1^{\bot}\t_2^{\bot}}\right)\, \ve_{2\t_2^{\bot}}\,\k(K_{2\t_2^{\bot}})
 +{\D p_{\t_2^{\bot}\t_{1^0}^{\bot}}\over 2}\,\k(K_{2\t_2^{\bot}}/2)  \, ,
\eea
 where each of the three lines corresponds respectively to the $\tau$, $\t_1^{\bot}$
 and $\t_{1^{\bot}}^{\bot}$ components and where now
 $\D p_{\t_2^{\bot}\t_{1^0}^{\bot}}\equiv |\langle \t_{1^0}^{\bot} | \t_2^{\bot} \rangle|^2 -
 |\langle  \bar{\t}_{1^0}^{\bot} | \bar{\t}_2^{\bot}  \rangle|^2 $.
 This last example shows, once more, how phantom terms are present whenever
 the production occurs either in one or in a two flavour regime, though only those generated
 by the heavier RH neutrinos can be afterwards asymmetrically washed out by the
 lighter RH neutrinos and contribute to the final asymmetry
 without cancelling with each other.

%%%%%%%%%%%%%%%
\section{Final discussion}
%%%%%%%%%%%%%%%

Within a Boltzmann classical kinetic formalism  one has
to distinguish the ten different RH neutrino mass patterns shown in Fig.~1.
These are obtained in the limits where the masses
$M_i$ are hierarchical and do not fall in the transition regimes.
We have extended the density matrix formalism for
the calculation of the matter-anti matter asymmetry in
leptogenesis including heavy neutrino flavours.
In this way we obtained  a  density matrix equation for the calculation of  
the asymmetry for any choice of the RH neutrino masses, even beyond the hierarchical limit.

Within this more general description,
the ten hierarchical RH neutrino mass patterns of Fig.~1 correspond
to those cases where the (five) different interactions are  only
one by one effective within a given range of temperatures.
In this way the evolution of the asymmetry can be described in well separated stages
where the density matrix equations greatly
simplify reducing to multiple sets of Boltzmann equations,
one for each stage.  In these cases we recovered or extended results that had already
been derived within a  simpler description based on an instantaneous collapse of lepton quantum  states.

The flavour projection effect,
where the orthogonal component of a previously produced asymmetry
escapes  the RH neutrino washout, is fully confirmed. We have
also shown that the washout of the parallel component is
exactly described by the usual exponential washout factor
independently of the washout regime.

Phantom terms emerge as quite a generic feature  of flavoured leptogenesis
and have to be taken into account even for vanishing initial RH neutrino abundances.
They can contribute to the final asymmetry even if the production from an heavier
RH neutrino species and the washout from a lighter RH neutrino species occur in the same
fully flavoured regime and so their presence goes beyond
the $N_2$-dominated scenario where they were originally discussed \cite{phantom}.
However, we have shown that, when the effect of gauge interactions in thermalising the
lepton abundances is taken into account, phantom terms get washed-out at the production,
though their wash-out rate is halved compared to that one acting on the final asymmetry.
In this way, in the strong wash-out regime, phantom terms give a contribution that is
also independent of the initial conditions.

Even though we have explicitly calculated the
final asymmetry only in one of the ten asymptotic limits RH neutrino mass
patterns shown in Fig.~1, in the case of the two RH neutrino model,  the procedure
can be  easily extended to all others neutrino mass patterns. For example
one can easily show the expression for the final asymmetry  in the
$N_2$-dominated scenario, when $M_1\ll 10^{9}\,{\rm GeV}$ \cite{geometry,vives,N2dom,phantom}.

It would be desirable in future to calculate the asymmetry beyond
these ten asymptotic limits, solving the full density matrix equation.
In this way  the  calculation of the matter-anti matter asymmetry  would
be extended  to a generic RH neutrino mass pattern, including the
cases where the  RH neutrino masses fall in the transition regimes
where quantum decoherence from charged lepton
interactions acts simultaneously with asymmetry generation and wash-out.
This would  make possible to interpolate between the asymptotic limits,
finding the exact conditions on the RH neutrino masses
for the validity of the solutions that we have discussed here.

\subsection*{Acknowledgments}

We wish to thank Stefan Antusch and Steve King for many useful discussions.
%PDB and DAJ wish to thank Bjorn Garbrecht for interesting
%discussions  during the Workshop on
%``Baryogenesis and First Order Phase Transitions in the Early Universe'',
%Lorentz Center,  29 August- 9 September,  Leiden, 2011.
PDB and LM acknowledge financial support from the NExT Institute and SEPnet.
SB acknowledges support from the Swiss National Science Foundation under
the Ambizione grant   PZ00P2\_136947. DAJ is thankful to the STFC for providing
studentship funding.

\section*{Appendix}
\appendix

\renewcommand{\thesection}{\Alph{section}}
\renewcommand{\thesubsection}{\Alph{section}.\arabic{subsection}}
\def\theequation{\Alph{section}.\arabic{equation}}
\renewcommand{\thetable}{\arabic{table}}
\renewcommand{\thefigure}{\arabic{figure}}
\setcounter{section}{1}
\setcounter{equation}{0}

In this Appendix we review and discuss some insightful aspects and properties of the heavy neutrino lepton
and anti-lepton bases, respectively $\{{\ell}_1,{\ell}_2,{\ell}_3\}$ and $\{\bar{\ell}_1,\bar{\ell}_2,\bar{\ell}_3\}$.

At tree level the two bases are $C\!P$ conjugated of each other and
 the probabilities $p^0_{ij} \equiv
|\langle {\ell_1^0} | {\ell_2^0} \rangle|^2 = |\langle {\bar{\ell}_1^0} | {\bar{\ell}_2^0} \rangle |^2$
can be expressed as \cite{problem}
\be
p^0_{ij}={\left|(m^{\dagger}_D\,m_D)_{ij}\right|^2
\over (m^{\dagger}_D\,m_D)_{ii}\,(m^{\dagger}_D\,m_D)_{jj}}
={|\sum_h\,m_h\,\O^{\star}_{hi}\,\O_{hj}|^2 \over \mti\,\mtj } \,  ,
\ee
where $\mti \equiv (m^{\dagger}_D\,m_D)_{ii}/M_i$.
In the last expression we expressed the terms $(m^{\dagger}_D\,m_D)_{ij}$
through the orthogonal matrix $\Omega$ providing a useful parameterisation of the
neutrino Dirac mass matrix given by
$m_D=U\,\sqrt{D_m}\,\Omega\,\sqrt{D_M}$ \cite{ci},
where $U$ is the leptonic mixing matrix, $D_M \equiv {\rm diag}(M_1,M_2,M_3)$
and $D_m \equiv {\rm diag}(m_1,m_2,m_3)$.

In general, the (tree level) bases $\{{\ell}^0_1,{\ell}^0_2,{\ell}^0_3\}$ and $\{\bar{\ell}^0_1,\bar{\ell}^0_2,\bar{\ell}^0_3\}$
are not orthonormal  (see Fig.~4) \cite{nardi}, i.e. in general $p^0_{ij}\neq \delta_{ij}$.
This case would indeed correspond to special forms of the Dirac mass matrices where
the orthogonal matrix is either the identity $(\Omega_{ij}=\delta_{ij})$,
or one of the other five special forms obtained from the identity  permuting
rows or columns. These six special forms imply \cite{geometry}
that the see-saw formula  reduces to the case where each light neutrino
mass $m_j$ proportional to a different inverse RH neutrino mass $M_i$,
so called form dominance models \cite{kingchen}.
However, when one of these six special cases are exactly realised and
the two bases are orthonormal,
both the total and the flavoured $C\!P$ asymmetries exactly vanish for the simple reason
that in this case  there is no interference between the tree level and the one loop graphs,
since this requires that in the decay of a RH neutrino $N_i$ a virtual RH neutrino $N_{j\neq i}$
couples to a lepton ${\ell}_i$ while orthonormality implies that it does not.

This means that for these six special forms, even including perturbative effects,
the heavy neutrino lepton and anti-lepton bases remain equal
to the  tree level bases and, therefore, they are still orthonormal.
Therefore, in order to have successful leptogenesis, the heavy neutrino
lepton and anti-lepton bases have  necessarily to be non-orthonormal to some level.

When some interference between tree level and one loop graphs is turned on,
implying non-orthonormality of the two bases, then
in general this will induce both non-vanishing total $C\!P$ asymmetries, with proportional
contributions in the flavoured $C\!P$ asymmetries given by the first terms in the eq.~(\ref{twofully}),
but also different flavour compositions between the heavy neutrino lepton basis
and heavy neutrino anti-lepton basis.  This can be seen easily from the expressions
for the flavoured $C\!P$ asymmetries recast in the orthogonal parameterisation
and for example in \cite{flavorlep} it was noticed how $\Delta p_{1\a}\neq 0$,
with a strong enhancement of the asymmetry compared to the unflavoured case (cf. eq.~(\ref{twofully}))
can be induced by the presence of low energy phases.
%%% I added this part, which is important I think to mention
As a matter of fact, it is well known
that the total $C\!P$ asymmetry depends only on a subset of phases (3) compared to the
flavoured $C\!P$ asymmetries. This difference can be precisely traced back to the
presence of the $\Delta p_{1\a}$ in Eq.~(\ref{twofully}), which includes the dependence
on the additional three phases.
%%%%
One can wonder why the flavour composition
of the final leptons and anti-leptons is affected by an account of the interference between tree
level and one loop graphs. In particular the neutrino Yukawa matrix can be always brought to
a triangular form $h = V^{\dagger}\,h_{\Delta}$, where $V$ is a unitary matrix. One can then switch
from the weak basis to another orthonormal
basis ${\ell}_{\Delta i} = V_{i\alpha}\,{\ell}_\alpha$. In this basis one has that at tree level
$N_1 \ra {\ell}_{\Delta 1} = {\ell}^0_1 = V_{1\a}{\ell}_{\a}$.  However,
this does not remain valid accounting for the interference with one loop graphs that make now
possible to have $N_1 \ra {\ell}_{j\neq 1}^0$. This clearly shows that, going beyond tree level,  the final ${\ell}_1$
is a linear combination of all three ${\ell}^0_i$, with a dominance of ${\ell}^0_1$
but also with a small contamination of ${\ell}^0_2$ and ${\ell}^0_3$. If one  considers
the anti-leptons there is also a deviation from $\bar{\ell}^0_1$ due to a contamination
of $\bar{\ell}^0_2$ and $\bar{\ell}^0_3$ that, however, is in general not $C\!P$
conjugated of the deviation in the ${\ell}_1$ from ${\ell}^0_1$.
At one loop these deviations are exactly described by the loop functions $\xi_u$ and $\xi_v$
in the eq. (\ref{uv}).
%%% New
Since we have that $\xi_u \neq \xi_v^{\star}$, it is clear from the
eqs.~(\ref{Cialpha})-(\ref{barCialpha}) that the the flavour compositions of the
lepton $\ell_1$ and of the antilepton $\bar{\ell}_1$ are different from each other,
as explicitly shown by the eq.~(\ref{deltapexplicit}).


\begin{thebibliography}{99}


\bibitem{fy}  M.~Fukugita and T.~Yanagida,
  %``Baryogenesis Without Grand Unification,''
  Phys.\ Lett.\ B {\bf 174}, 45 (1986).


\bibitem{seesaw} P.~Minkowski,
Phys.\ Lett.\ B {\bf 67},  421 (1977);
M. Gell-Mann, P. Ramond and
R. Slansky,  {\em Proceedings of the Supergravity Stony Brook Workshop}, New
York 1979,  eds. P. Van Nieuwenhuizen and D. Freedman; T. Yanagida,  {\em
Proceedings of the Workshop on Unified Theories and Baryon Number in the
Universe},  Tsukuba, Japan 1979, ed.s A. Sawada and A. Sugamoto;
R. N. Mohapatra, G. Senjanovic, Phys. Rev. Lett. {\bf 44},  912 (1980).

\bibitem{neuosc}
 Y.~Fukuda {\it et al.} [ Super-Kamiokande Collaboration ],
  %``Evidence for oscillation of atmospheric neutrinos,''
  Phys.\ Rev.\ Lett.\  {\bf 81 } (1998)  1562-1567 [hep-ex/9807003].

\bibitem{DL2}
A.~D.~Dolgov and Ya.~B.~Zeldovich, Rev. Mod. Phys. {\bf 53} (1981) 1;
E.~W.~Kolb and S.~Wolfram, Nucl. Phys. B{\bf 172} (1980) 224, ibid. {\bf B 195} (1982) 542 (E).

\bibitem{early}
M.~A.~Luty, %``Baryogenesis via leptogenesis,''
Phys.\ Rev.\  {\bf D45 } (1992)  455-465;
M.~Plumacher,
  %``Baryogenesis and lepton number violation,''
  Z.\ Phys.\  {\bf C74 } (1997)  549-559;
 W.~Buchmuller, M.~Plumacher,
  %``Neutrino masses and the baryon asymmetry,''
  Int.\ J.\ Mod.\ Phys.\  {\bf A15 } (2000)  5047-5086.



\bibitem{bcst}
R.~Barbieri, P.~Creminelli, A.~Strumia, N.~Tetradis,
  %``Baryogenesis through leptogenesis,''
  Nucl.\ Phys.\  {\bf B575 } (2000)  61-77.

\bibitem{cmb}
W.~Buchmuller, P.~Di Bari and M.~Plumacher,
  %``Cosmic microwave background, matter - antimatter asymmetry and neutrino masses,''
  Nucl.\ Phys.\ B {\bf 643} (2002) 367 [Erratum-ibid.\ B {\bf 793} (2008) 362].
  %%CITATION = HEP-PH/0205349;%%

\bibitem{giudice}
G.~F.~Giudice, A.~Notari, M.~Raidal, A.~Riotto and A.~Strumia, Nucl.\ Phys.\  B {\bf 685} (2004) 89.

 \bibitem{pedestrians}
 W.~Buchmuller, P.~Di Bari, M.~Plumacher,
  %``Leptogenesis for pedestrians,''
  Annals Phys.\  {\bf 315 } (2005)  305-351.


\bibitem{nardi}
E.~Nardi, Y.~Nir, E.~Roulet, J.~Racker,
  %``The Importance of flavor in leptogenesis,''
  JHEP {\bf 0601 } (2006)  164.
  [hep-ph/0601084].

\bibitem{abada}
  A.~Abada, S.~Davidson, F.~-X.~Josse-Michaux, M.~Losada, A.~Riotto,
  %``Flavor issues in leptogenesis,''
  JCAP {\bf 0604}, 004 (2006).
  [hep-ph/0601083].

\bibitem{flavorlep}
 S.~Blanchet, P.~Di Bari,
  %``Flavor effects on leptogenesis predictions,''
  JCAP {\bf 0703 } (2007)  018.
  [hep-ph/0607330].


\bibitem{geometry}
P.~Di Bari,
 %``Seesaw geometry and leptogenesis,''
 Nucl.\ Phys.\  {\bf B727 } (2005)  318-354.
  [hep-ph/0502082].

\bibitem{vives}
 O.~Vives,
  %``Flavor dependence of $C\!P$ asymmetries and thermal leptogenesis with strong right-handed neutrino mass hierarchy,''
  Phys.\ Rev.\  {\bf D73 } (2006)  073006.
  [hep-ph/0512160].



\bibitem{2RHneutrino}
  S.~Antusch, P.~Di Bari, D.~A.~Jones, S.~F.~King,
  %``Leptogenesis in the two right-handed neutrino model revisited,''
 Phys. \ Rev. {\bf D 86}, (2012) 023516 [arXiv:1107.6002 [hep-ph]].

\bibitem{SO10lep}
P.~Di Bari, A.~Riotto,
  %``Successful type I Leptogenesis with SO(10)-inspired mass relations,''
  Phys.\ Lett.\  {\bf B671 } (2009)  462-469;
  %``Testing SO(10)-inspired leptogenesis with low energy neutrino experiments,''
  JCAP {\bf 1104 } (2011)  037.

\bibitem{problem}
  E.~Bertuzzo, P.~Di Bari, L.~Marzola,
  %``The problem of the initial conditions in flavoured leptogenesis and the tauon N_2-dominated scenario,''
  Nucl.\ Phys.\  {\bf B849 } (2011)  521-548.

 \bibitem{phantom}
 S.~Antusch, P.~Di Bari, D.~A.~Jones and S.~F.~King,
  %``A fuller flavour treatment of N_2-dominated leptogenesis,''
  Nucl.\ Phys.\ B {\bf 856} (2012) 180.
  %%CITATION = ARXIV:1003.5132;%%

 \bibitem{riottodesimone}
A.~De Simone, A.~Riotto,
  %``On the impact of flavour oscillations in leptogenesis,''
  JCAP {\bf 0702 } (2007)  005.
  [hep-ph/0611357].

 \bibitem{nir}
G.~Engelhard, Y.~Grossman, E.~Nardi, Y.~Nir,
  %``The Importance of N2 leptogenesis,''
  Phys.\ Rev.\ Lett.\  {\bf 99 } (2007)  081802.






\bibitem{momentum}
A.~Basboll and S.~Hannestad, JCAP {\bf 0701} (2007) 003;
F.~Hahn-Woernle, M.~Plumacher and Y.~Y.~Y.~Wong, JCAP {\bf 0908} (2009) 028.

\bibitem{thermal}
C.~P.~Kiessig, M.~Plumacher and M.~H.~Thoma, Phys.\ Rev.\  D {\bf 82} (2010) 036007.

\bibitem{spectators}
W.~Buchmuller and M.~Plumacher, Phys.\ Lett.\ B {\bf 511} (2001) 74.
  %%CITATION = HEP-PH/0104189;%%

\bibitem{N2dom}
S.~Blanchet and P.~Di Bari,
  %``New aspects of leptogenesis bounds,''
  Nucl.\ Phys.\ B {\bf 807} (2009) 155.
  %%CITATION = ARXIV:0807.0743;%%


\bibitem{qke}
A.~De Simone and A.~Riotto, JCAP {\bf 0708} (2007) 002;
M.~Beneke, B.~Garbrecht, M.~Herranen and P.~Schwaller, Nucl.\ Phys.\  B {\bf 838} (2010) 1;
A.~Anisimov, W.~Buchmuller, M.~Drewes and S.~Mendizabal, arXiv:1012.5821.

\bibitem{WMAP7}
  E.~Komatsu {\it et al.} [ WMAP Collaboration ],
  %``Seven-Year Wilkinson Microwave Anisotropy Probe (WMAP) Observations: Cosmological Interpretation,''
  Astrophys.\ J.\ Suppl.\  {\bf 192 } (2011)  18.
  [arXiv:1001.4538 [astro-ph.CO]].

\bibitem{beneke}
M.~Beneke, B.~Garbrecht, C.~Fidler, M.~Herranen, P.~Schwaller,
  %``Flavoured Leptogenesis in the CTP Formalism,''
  Nucl.\ Phys.\  {\bf B843 } (2011)  177-212.
  [arXiv:1007.4783 [hep-ph]].


\bibitem{di}
S.~Davidson, A.~Ibarra,
  %``A Lower bound on the right-handed neutrino mass from leptogenesis,''
  Phys.\ Lett.\  {\bf B535 } (2002)  25-32.

\bibitem{zeno}
 S.~Blanchet, P.~Di Bari, G.~G.~Raffelt,
  %``Quantum Zeno effect and the impact of flavor in leptogenesis,''
  JCAP {\bf 0703 } (2007)  012.
  [hep-ph/0611337].

\bibitem{feynman}
Feynman, Richard P. , ``Statistical Mechanics: A Set of Lectures", Addison Wesley (1981).

%\cite{Buchmuller:1997yu}
\bibitem{Buchmuller:1997yu}
  W.~Buchmuller, M.~Plumacher,
  %``CP asymmetry in Majorana neutrino decays,''
  Phys.\ Lett.\  {\bf B431}, 354-362 (1998).
  [hep-ph/9710460].

%\cite{Anisimov:2005hr}
\bibitem{Anisimov:2005hr}
  A.~Anisimov, A.~Broncano, M.~Plumacher,
  %``The CP-asymmetry in resonant leptogenesis,''
  Nucl.\ Phys.\  {\bf B737}, 176-189 (2006).

%\cite{Covi:1996wh}
\bibitem{cr}
  L.~Covi, E.~Roulet and F.~Vissani,
  %``CP violating decays in leptogenesis scenarios,''
  Phys.\ Lett.\ B {\bf 384}, 169 (1996)
  [hep-ph/9605319].
  %%CITATION = HEP-PH/9605319;%%

\bibitem{sigl}
G.~Sigl and G.~Raffelt,
  %``General kinetic description of relativistic mixed neutrinos,''
  Nucl.\ Phys.\ B {\bf 406} (1993) 423.
  %%CITATION = NUPHA,B406,423;%%

\bibitem{thesis}
  S.~Blanchet, ``A New Era of Leptogenesis'', Ph.D thesis,  [arXiv:0807.1408 [hep-ph]].


\bibitem{weldon}
  H.~A.~Weldon,
  %``Effective Fermion Masses of Order gT in High Temperature Gauge Theories with Exact Chiral Invariance,''
  Phys.\ Rev.\  {\bf D26}, 2789 (1982).

\bibitem{cline}
  J.~M.~Cline, K.~Kainulainen, K.~A.~Olive,
  %``Protecting the primordial baryon asymmetry from erasure by sphalerons,''
  Phys.\ Rev.\  {\bf D49}, 6394-6409 (1994).

\bibitem{raffelt}
A.~D.~Dolgov, S.~H.~Hansen, S.~Pastor, S.~T.~Petcov, G.~G.~Raffelt and D.~V.~Semikoz,
  %``Cosmological bounds on neutrino degeneracy improved by flavor oscillations,''
  Nucl.\ Phys.\ B {\bf 632} (2002) 363.

\bibitem{activesterile}
R.~Foot and R.~R.~Volkas,
  %``Reconciling sterile neutrinos with big bang nucleosynthesis,''
  Phys.\ Rev.\ Lett.\  {\bf 75} (1995) 4350;
  %%CITATION = HEP-PH/9508275;%%
  P.~Di Bari, P.~Lipari and M.~Lusignoli,
  %``The muon-neutrino <---> s neutrino interpretation of the atmospheric neutrino data and cosmological constraints,''
  Int.\ J.\ Mod.\ Phys.\ A {\bf 15} (2000) 2289;
  %%CITATION = HEP-PH/9907548;%%
 D.~Kirilova,
  %``BBN with Late Electron-Sterile Neutrino Oscillations: The Finest Leptometer,''
  JCAP {\bf 1206} (2012) 007.
  %%CITATION = ARXIV:1101.4177;%%

\bibitem{hierarchical}
S.~Blanchet, P.~Di Bari,
  %``Leptogenesis beyond the limit of hierarchical heavy neutrino masses,''
  JCAP {\bf 0606 } (2006)  023.

\bibitem{resonant}
A.~Pilaftsis and T.~E.~J.~Underwood,
  %``Resonant leptogenesis,''
  Nucl.\ Phys.\ B {\bf 692} (2004) 303
  [hep-ph/0309342].
  %%CITATION = HEP-PH/0309342;%%

\bibitem{subtleresonant}
A.~De Simone and A.~Riotto, %``On Resonant Leptogenesis,''
JCAP {\bf 0708} (2007) 013 [arXiv:0705.2183 [hep-ph]].
  %%CITATION = ARXIV:0705.2183;%%

\bibitem{fgy}
S.~F.~King,
  %``Large mixing angle MSW and atmospheric neutrinos from single right-handed neutrino dominance and U(1) family symmetry,''
  Nucl.\ Phys.\ B {\bf 576} (2000) 85;
  %%CITATION = HEP-PH/9912492;%%
P.~H.~Frampton, S.~L.~Glashow and T.~Yanagida,
  %``Cosmological sign of neutrino CP violation,''
  Phys.\ Lett.\  B {\bf 548} (2002) 119.
  %%CITATION = PHLTA,B548,119;%%

 %\cite{King:2002qh}
\bibitem{various}
  S.~F.~King,
  %``Leptogenesis - MNS link in unified models with natural neutrino mass
  %hierarchy,''
  Phys.\ Rev.\  D {\bf 67} (2003) 113010 [arXiv:hep-ph/0211228];
  %%CITATION = PHRVA,D67,113010;%%
  P.~H.~Chankowski and K.~Turzynski,
  %``Limits on T(reh) for thermal leptogenesis with hierarchical neutrino
  %masses,''
  Phys.\ Lett.\  B {\bf 570} (2003) 198 [arXiv:hep-ph/0306059];
  %%CITATION = PHLTA,B570,198;%%
 A.~Ibarra and G.~G.~Ross,
  %``Neutrino phenomenology: The case of two right handed neutrinos,''
  Phys.\ Lett.\  B {\bf 591} (2004) 285
  [arXiv:hep-ph/0312138].
  %%CITATION = PHLTA,B591,285;%%

\bibitem{ci}
J.~A.~Casas and A.~Ibarra,
  %``Oscillating neutrinos and muon ---> e, gamma,''
  Nucl.\ Phys.\ B {\bf 618} (2001) 171.
  %%CITATION = HEP-PH/0103065;%%

\bibitem{kingchen}
M.~C.~Chen and S.~F.~King,
  %``A4 See-Saw Models and Form Dominance,''
  JHEP {\bf 0906} (2009) 072
  [arXiv:0903.0125 [hep-ph]].
  %%CITATION = JHEPA,0906,072;%%

\end{thebibliography}
\end{document}